# Generalized formulation for ideal light-powered systems through energy and entropy flow analysis

## Part1. Based on the first-order evaluation

Tetsuo Yabuki

School of Economics, Hokusei Gakuen University. 2-3-1 Oyachi-nishi, Atsubetsu-ku,

Sapporo,004-8631, Japan

E-mail address: t-yabuki@hokusei.ac.jp

**Abstract** (250words)

In this study, the theoretical maximum efficiency $\eta_{max}$ and the Boltzmann-type factor $[P^*]/\ [P]$ giving the concentration ratio of excited-to-ground state pigment-molecules for photosynthetic systems under irradiation with arbitrary photon flux density $n_\gamma(\lambda)$, solid angle $\Omega$, and degree of polarization $P$, are formulated in the most fundamental and general way through energy and entropy flow analysis, using reversibility and the first-order evaluable condition by the photon number change, which is a quasi-equilibrium condition between the radiation and the system, as essential conditions. The radiation temperature for the diluted monochromatic light as non-equilibrium, obtained by the fundamental formulation of this study is found to agree with the conventional radiation temperature, often called the effective temperature, for a given photon flux density (light intensity), provided that $\Omega$ and $P$ of the radiation are $4\pi$ and 0, respectively. The reason for this agreement is discussed in the final section. The formulation in this study allows quantitative analyses that are not possible with conventional radiation temperature. As examples, the formulation of $\eta_{max}$ taking into account entropy changes due to photochemical reactions such as glucose production by photosynthesis, and various quantities under irradiation at arbitrary $\Omega$ and $P$ are presented. In Appendix C, a specific and rigorous proof, using elementary geometry, of the fact that the entropy of radiation diluted on its way from the Sun to the Earth remains unchanged from its original value, until it is scattered by the Earth's atmosphere, which is guaranteed only in general terms by Liouville's theorem, is given.

## 1. Introduction

Solar energy and the natural and artificial light-powered systems using it have attracted substantial consideration from the perspectives of agriculture and global warming problems, and their theoretical and actual maximum efficiencies have been vigorously discussed [1–4]. However, despite extensive research on the theoretical calculation of the theoretical maximum energy efficiency, i.e. the ideal efficiency for monochromatic light and non-monochromatic light (for the former see the first paper [5], and for details of the latter, see Part 2 of this study [6]), certain questions remain unanswered, especially around temperature and the second law of thermodynamics [7-14]. These unresolved questions have been considered in the present study on two levels. One is the problem within the range of the first-order evaluable condition for monochromatic light and is analyzed in this Part1. The other is the problem when blackbody radiation is considered beyond the first-order evaluable condition and is analyzed in Part2 of this study.

The ideal efficiency of photosynthesis at the surface of the Earth was first formulated by Duysens [5] in the 1950s as a theoretical maximum energy efficiency. He considered the dilution effect of sunlight emitted by the Sun as blackbody radiation at about 5800 K, reaching the surface of the Earth from a distance of about 150 million km. His formulation involved two steps: 1) Substituting the photon number flux (light intensity) reduced by the dilution effect into the blackbody radiation equation (Planck radiation equation) and solving for temperature; 2) Substituting the resulting temperature (called the effective temperature) into Carnot efficiency formula, the ideal efficiency of a heat engine. However, owing to dilution effects, the sunlight reaching the surface of the Earth does not remain blackbody radiation, making this effective temperature wavelength dependent, represented as $T_D(\lambda)$ in this paper. Although this effective temperature has been widely used in the physicochemical analysis of the theoretical efficiency of photosynthesis, certain questions and critical discussions remain unresolved [7-14]. These questions stem mainly from the fact that this temperature $T_D(\lambda)$, which is not in equilibrium (i.e., not a true temperature), is automatically applied to the expressions for Carnot efficiency, and the Boltzmann-type factor formula giving the ratio of excited-state to grand-state pigment concentrations, as $\eta_{max} = 1 - T_{out}/T_D(\lambda)$ and $\exp(-\Delta E/k_B T_D(\lambda))$, respectively, without proper justification. This approach is not only logically flawed but also fails to accurately calculate ideal efficiencies that consider various variables, such as solid angle, polarization, absorption rate (the number ratio of photons absorbed by the system to photons irradiated), and entropy changes due to photochemical reactions within the system.

For example, it has been reported that the energy efficiency $\eta$ of photosynthesis obtained in an experiment using artificial light such as lasers exceeds the ideal efficiency obtained by this effective radiation temperature $T_D(\lambda)$, which is converted from photon flux density (light intensity) [14]. It has been commented that thermodynamic principles might not be applicable to photonic systems such as photosynthetic photosystems, but correctly, this is a result of the low entropy property of laser light due to its small solid angle $\Omega$ which could not be properly evaluated at the conventional effective temperature. The resulting value of the energy efficiency $\eta$ obtained in the experiment is consistent with the estimated value at the radiation temperature formulated by considering the solid angle of the laser light in this study. A detailed quantitative analysis of the results is presented in Section 5.3.

Furthermore, this method implicitly assumes a quasi-equilibrium condition between the radiation and system, which precludes the analysis of non-equilibrium contributions between them, as addressed in Part2 of this study. In contrast, the method used in this study was first to derive a fundamental expression for the theoretical maximum energy efficiency $\eta_{max}$ by energy and entropy flow analysis based on the second law of thermodynamics, and then to extract the temperature-dimensional quantity from an expression of $\eta_{max}$. This quantity coincides with the conventional radiation temperature $T_D(\lambda)$, called the effective temperature by Duysens [5], within the first-order evaluation by photon number variation, provided that the solid angle $\Omega$ and the degree of polarization $P$ of the light are $4\pi$ and 0, respectively. The reason for this coincidence is discussed in Section 8.3. Thus, in Part 1 of this study, contradictions and controversies that have been pointed out and still exist today were resolved within the framework of first-order evaluation. The resolution in a framework beyond the first-order evaluation is done in Part 2 of this study.

In this study, the term "light-powered system" is defined not only as a system that outputs electrical energy, such as solar power generation, photovoltaics and solar cells, etc., but also as a system from which any kind of physical work is extracted using light energy, including natural and artificial photosynthesis.

The formulation used in this study is so general that it has been possible to formulate a number of formulae that were not possible using the conventional methods described above. The main ones from the formulations in this study are listed below.

1. This formulation allows for formula that correctly incorporate arbitrary degree of polarization. Specifically, its ideal efficiency is shown in Eq (4.17) and its radiation temperature is derived from it in Eq (4.18). The results of the calculation using these equations are shown in Fig. 3 and Fig. 5.
2. In this formulation, the ideal efficiency $\eta_{max}$ of the photosynthesis system on the Earth incorporating entropy changes due to photochemical reaction was formulated as shown in Eq. (6.7). The results of using this equation to calculate the ideal efficiency using actual absorption spectra in PAR (Photosynthetic Active Radiation) of several plants [15] are shown in the table.
3. In this formulation, by considering the photochemical transduction from radiation to a two-level pigment system in terms of entropy, it became possible to correctly derive the general formula for the concentration ratio of the pigment in the excited state to that in the ground state, as shown in Eq. (7.15), including an arbitrary solid angle $\Omega$ and polarization parameter $\alpha_1, \alpha_2$ of the irradiated light. When $\Omega = 4\pi$ and $\alpha_1 = \alpha_2 = 1/2$ (completely unpolarized), the result was consistent with the conventional Boltzmann-type factor formula using the effective temperature.

The results of the calculation using equation (7.15) are shown in Figure 6.

Finally, it was shown that the entropy of the radiation diluted on its way from the Sun to the Earth does not increase until it is scattered by the Earth's atmosphere, and remains unchanged from its original value. This is because the increase in entropy in configuration space is completely canceled out by the decrease in entropy in momentum-space due to the decrease in solid angle $\Omega$ which gives the momentum directional degrees of freedom. A specific and rigorous proof of this fact, which is guaranteed only in general terms by Liouville's theorem, is given using elementary geometry in Appendix C. Therefore, the following points should be noted regarding the formulation of the ideal efficiency of photosynthesis under diluted sunlight from the Sun to the Earth's surface using the effective temperature in previous studies. The decrease in entropy in momentum-space is eliminated by the recovery of the solid angle due to scattering in the Earth's atmosphere, and the increase in entropy due to the dilution effect and the associated decrease in the ideal efficiency of the light engine becomes apparent. When performing such an analysis in a real field cite, it is necessary to take into account the wavelength dependence of scattering on the Earth.
The analysis in this study is based on a particle picture in which the quantum coherence of photons is zero, assuming that the light source is a blackbody radiator, such as sunlight. And the quantum coherence of the photons emitted from the laser light source is not taken into account.
In this paper, the term "light-powered system " is used to refer to natural photosynthesis, artificial photosynthesis, photovoltaics and solar cells, and their “theoretical maximum energy efficiency” and “ideal efficiency” are used interchangeably. In addition, “light intensity” is used as an interchangeable term for photon flux density. Also, from Section 4 onwards, quantities related to radiation will be marked with the subscript $\gamma$, e.g. $\eta_{max}^{\gamma}(\lambda)$, $E_{in}^{\gamma}(\lambda)$, $S_{in}^{\gamma}(\lambda)$, $N_{in}^{\gamma}(\lambda)$, etc.

## 2. General formulation of radiation entropy

The entropy of radiation is defined by applying the mathematical formula, $S = k_B \ln W$ originally formulated by Boltzmann in statistical mechanics, to radiation as a population of photons, where $k_B$ is the Boltzmann constant equal to 1.38 × $10^{-23}$ [J/K], ln is the natural logarithm, and $W$ is the total number of microscopic accessible states for a particle constituting the macroscopic state of an ensemble.

### 2.1. Monochromatic light with wavelength $\lambda$ and solid angle $\Omega$, without polarization

This subsection presents a general mathematical formulation of the entropy $S(\lambda, \Omega)$ of monochromatic radiation with wavelength $\lambda$ and arbitrary solid angle $\Omega$, without polarization, based on the above formula. Based on the fact that the photon is a Bose particle, the starting equation is the following formula for the quantum statistical entropy of a Bose particle ensemble, which is valid for thermodynamic states including non-equilibrium:

$$S(\lambda,\Omega) = k_B G(\lambda,\Omega)\{(1+f(\lambda,\Omega))\ln(1+f(\lambda,\Omega)) - f(\lambda,\Omega)\ln f(\lambda,\Omega)\}, \qquad (2.1)$$

which can be derived from $W = {}_{N+G-1}C_N = \frac{(N+G-1)!}{(G-1)!N!}$, i.e., the number of cases in which $N$ indistinguishable Bose particles can be arranged in $G$ distinguishable quantum states, based on the Stirling approximation. In this study $N = N(\lambda,\Omega)$ and $G = G(\lambda,\Omega)$ are the photon number and the number of quantum states, of a photon with wavelength $\lambda$ and solid angle $\Omega$, respectively. And $f(\lambda,\Omega)$ is a distribution function expressed as $f(\lambda,\Omega) = \frac{N(\lambda,\Omega)}{G(\lambda,\Omega)}$. In general, the number of quantum states $G$ contained in a phase space volume $\Delta q^3 \Delta p^3$ can be counted, considering the unit $h^3$ (where $h = 6.63 \times 10^{-34}$[Js] is the Planck constant), based on the uncertainty relation between $q$ and $p$, $\Delta q \Delta p \sim h$,as follows:

$$G = \frac{\Delta q^3 \Delta p^3}{h^3}, \qquad (2.2)$$

where $\Delta q^3$ is the volume of configuration space for particles to be able to access. Thus, it is generally identified as $\Delta q^3 = V$, where $V$ is the volume occupied by particles. When the magnitude of the momentum of particles is within a sufficiently small range between $p$ and $p + \Delta p$, and the direction of the momentum is within a solid angle $\Omega$, $\Delta p^3$ can be expressed as $\Delta p^3 = \Omega p^2 \Delta p$, which is the volume of momentum accessible to the particle. Consequently, we obtain $G(p,\Omega) = \frac{V\Omega p^2 \Delta p}{h^3}$ and $f(p,\Omega) = \frac{N(p,\Omega)}{G(p,\Omega)} = \frac{\rho_N(p,\Omega)h^3}{\Omega p^2 \Delta p}$, where $\rho_N(p,\Omega) = N(p,\Omega)/V$ represents the number density of the particles. Since the momentum of a photon is expressed as $p = h/\lambda$ in relation to the wavelength $\lambda$ of the classical wave picture, a sufficiently small range $\Delta p$ can be expressed as $\Delta p = (h/\lambda^2)\Delta\lambda$. Considering that a photon is a massless particle with spin $s = 1$, it has only two spin z components, $s_z = -1$ and 1, which is associated with the polarization of light as an electromagnetic wave. Thus, after multiplying **Eq. (2.2)** by a numerical factor of 2, corresponding to this degree of freedom of a photon's spin, the following formula is obtained for the quantum state formula $G(\lambda,\Omega)$ and the distribution function $f(\lambda,\Omega)$, respectively:

$$G(\lambda,\Omega) = \frac{2\Omega V}{\lambda^4}\Delta\lambda, \qquad f(\lambda,\Omega) = \frac{\rho_N(\lambda)\,\lambda^4}{2\Omega\Delta\lambda}. \qquad (2.3)$$

In the case of thermal equilibrium at temperature $T$, i.e., blackbody radiation, $f_{BB}(\lambda) = \frac{1}{\exp\left(\frac{hc}{\lambda k_B T}\right)-1}$ is obtained as the Bose-Einstein distribution function, which maximizes the entropy expressed in Eq. (2.1) under the constraint condition that the total energy is constant, where $c$ is the velocity of light. Since thermal radiation (blackbody radiation) is an isotropic (the direction of the momentum is uniformly distributed in the solid angle $\Omega = 4\pi$) and unpolarized light (the degree of polarization $P = 0$) ensemble, and the photon flux density per wavelength $\hat{n}(\lambda)$ is given by $(1/4)\rho_N(\lambda, T)c/\Delta\lambda$, the following Planck's spectral distribution law for the number flux per wavelength of blackbody radiation, $\hat{n}_{BB}(\lambda) = \frac{2\pi c}{\lambda^4}\frac{1}{\exp\left(\frac{hc}{\lambda k_B T}\right)-1}$, is obtained. ,

### 2.2. Thermodynamic formulation of the radiation entropy including the degree of polarization

When analyzing photosynthesis on Earth, we must consider all forms of scattering of sunlight, such as reflection, transmission, and scattering by small particles, especially in the atmosphere and clouds, which inevitably result in polarization $P$. So far, several formulations for the entropy of radiation with an arbitrary $P$ have been proposed; however, most of these entropies are only on the measure for the degree of polarization [16–20], but not the exact thermodynamic entropy, which can analyze physical quantities such as the energy efficiency of light-powered systems. There have been several studies of thermodynamic polarization entropy [21-23], but almost all of them are basically rooted in Shannon's information entropy as a measure of polarizability, and no precise, general form has yet been presented, except for the case of perfectly polarized or unpolarized light. Therefore, in this subsection, the author's most general formulation of the entropy of radiation with arbitrary polarization, based on the discussion in the previous subsection, is summarized below.

The polarization of an electromagnetic wave in the classical picture of radiation corresponds to the z-component (±1) of the spin of the photon in the quantum picture. From this consideration, the entropy, which includes the polarization, is expressed as follows.

$$S(\lambda, \Omega) = \sum_{i=1}^{2} S(\lambda, \Omega, \alpha_i), \tag{2.4}$$

where $\alpha_i (\, i = 1,2 : 0 \leq \alpha_1, \alpha_2 \leq 1, \alpha_1 + \alpha_2 = 1)$ are the relative ratios of the number of photons in $s_z = \pm 1$. They are the relative ratios of the eigenvalues of the coherence matrix, known in classical optics as the polarization matrix, and are related to the degree of polarization $P$ in the form $|\alpha_1 - \alpha_2| = P$ from the definition of $P$.

Furthermore, $S(\lambda, \Omega, \alpha_i)$ is expressed as

$$S(\lambda, \Omega, \alpha_i) = k_B \frac{G(\lambda,\Omega)}{2}\left\{\left(1 + f(\lambda, \Omega, \alpha_i)\right)\ln\left(1 + f(\lambda, \Omega, \alpha_i)\right) - f(\lambda, \Omega, \alpha_i)\ln f(\lambda, \Omega, \alpha_i)\right\}, \tag{2.5}$$

where $f(\lambda, \Omega, \alpha_i)$ is expressed as the following formula according to the definition $f \equiv \frac{N}{G}$,

$$f(\lambda,\Omega,\alpha_i)=\frac{N(\lambda,\Omega,\alpha_i)}{G_i(\lambda,\Omega)}=\frac{\alpha_i N(\lambda,\Omega)}{(1/2)G(\lambda,\Omega)}=2\alpha_i f(\lambda,\Omega). \quad (2.6)$$

Using Eqs. (2.5)- (2.6), the following is obtained:

$$S(\lambda,\Omega,\alpha_1,\alpha_2)\ =k_B\frac{G(\lambda,\Omega)}{2}\sum_{i=1}^{2}\{\big(1+2\alpha_i f(\lambda,\Omega)\big)\ln\big(1+2\alpha_i f(\lambda,\Omega)\big)-f(\lambda,\Omega)\ln f(\lambda,\Omega)\}$$

$$-k_B N(\lambda,\Omega)\big(\ln 2+\sum_{i=1}^{2}\alpha_i\ln\alpha_i\big). \quad (2.7)$$

In the case of unpolarized light, such as sunlight, with $\alpha_1=\alpha_2=1/2$, Eq. (2.7) is an overwhelming maximum and consequently reduces to the general formula in Eq. (2.1). Furthermore, in the case of completely polarized light, specifically, $\alpha_1=0,\alpha_2=1$ or $\alpha_1=1,\alpha_2=0$, the entropy of radiation is minimized and reduces to the complete polarization entropy

$$S_{CP}(\lambda,\Omega)$$

$$=k_B\frac{G(\lambda,\Omega)}{2}\{\big(1+2f(\lambda,\Omega)\big)\ln\big(1+2f(\lambda,\Omega)\big)-2f(\lambda,\Omega)\ln\big(2f(\lambda,\Omega)\big)\}. \quad (2.8)$$

Certain studies [20-23] have provided formulae for the radiation entropy including its degree of polarization P; however, they differ subtly but essentially from Eq. (2.7), almost all of which presented a formula only for the completely polarized state.

When $f(\lambda,\Omega)=\frac{N(\lambda)}{G(\lambda,\Omega)}\ll 1$, Eq. (2.7) and Eq. (2.8) are reduced to

$$S(\lambda,\Omega,\alpha_1,\alpha_2)$$

$$=k_B G(\lambda,\Omega)\{f(\lambda,\Omega)-f(\lambda,\Omega)\ln f(\lambda,\Omega)\}-k_B N(\lambda,\Omega)\big(\ln 2+\sum_{i=1}^{2}\alpha_i\ln\alpha_i\big), \quad (2.9)$$

$$S_{CP}(\lambda,\Omega)=k_B G(\lambda,\Omega)\{f(\lambda,\Omega)-f(\lambda,\Omega)\ln f(\lambda,\Omega)\}-k_B N(\lambda,\Omega)\ln 2. \quad (2.10)$$

respectively. From Eqs. (2.9) and (2.10), when we evaluate the entropy using $S_{CP}(\lambda,\Omega)$ as the reference value, $\tilde{S}(\lambda,\Omega)\equiv S(\lambda,\Omega)-S_{CP}(\lambda,\Omega)$, then we have

$$\tilde{S}(\lambda,\Omega)=-k_B N\sum_{i=1}^{2}\alpha_i\ln\alpha_i. \quad (2.11)$$

This term has the form of Shannon entropy, and it can be represented by the degree of polarization $P$ as follows:

$$-\sum_{i=1}^{2}\alpha_i ln\alpha_i=-\left\{\frac{1-P}{2}\ln\left(\frac{1-P}{2}\right)+\frac{1+P}{2}\ln\left(\frac{1+P}{2}\right)\right\}, \quad (2.12)$$

where the equality $P=|\alpha_1-\alpha_2|$ is used. Here, Eq. (2.12) is the so-called polarization entropy in optical science [16-20]. From the above discussion, we can see that $-N\sum_{i=1}^{2}\alpha_i\ln\alpha_i$ deserves a thermodynamic measure of polarization entropy only under the condition that $N(\lambda,\Omega)/G(\lambda,\Omega)\ll 1$ is satisfied. In the formulation of the dependence of the ideal efficiency $\eta_{max}$ of light-powered systems on the degree of polarization in Section 4.3.4, a cleaner and more rigorous formulation is used without this approximation. From the above discussion, we can see that $\big(-\sum_{i=1}^{2}\alpha_i ln\alpha_i\big)$ simply represents the information entropy that measures the degree of polarization (the entropy at complete polarization is 0), and is not the thermo-statistical entropy that is useful in thermodynamics, to which the polarization degree contributes. In this study, the thermodynamic entropy (not just the information entropy) of the degree of polarization is precisely incorporated into the ideal efficiency (Section 4) and the concentration ratio of excited-to-ground state pigment-molecules (the Boltzmann-type factor) (Section 7), directly using Eq. (2.6) (not using the approximation condition $f(\lambda, \Omega) \ll 1$) within the framework of the first-order evaluation.

## 3. Entropy flow accompanying energy flow and ideal efficiency in a general powered system

The general powered system, including photosynthesis, consists of three parts: an energy source, a system that outputs work, and a heat sink (environment outside the system). The energy flow consists of three flows: the flow $E_{in}$ from the energy source to the system, the flow $W_{out}$ extracted as work and the flow $Q_{out}$ emitted as waste heat required for entropy discard.

The entropy flow associated with the energy flow satisfies the following equation (Fig. 1):

$$S_{out} = S_{in} + S_g - S_W, \tag{3.1}$$

where the four types of entropy flow in Eq. (3.1) are represented as follows.

$$\begin{aligned} &S_{in}: \text{Entropy imported into a powered system by the source energy} \\ &S_g: \text{Entropy generated in a powered system} \\ &S_{out}: \text{Entropy wastefully emitted by a powered system} \\ &S_W: \text{Entropy contained in the power outputted by a powered system} \end{aligned} \tag{3.2}$$

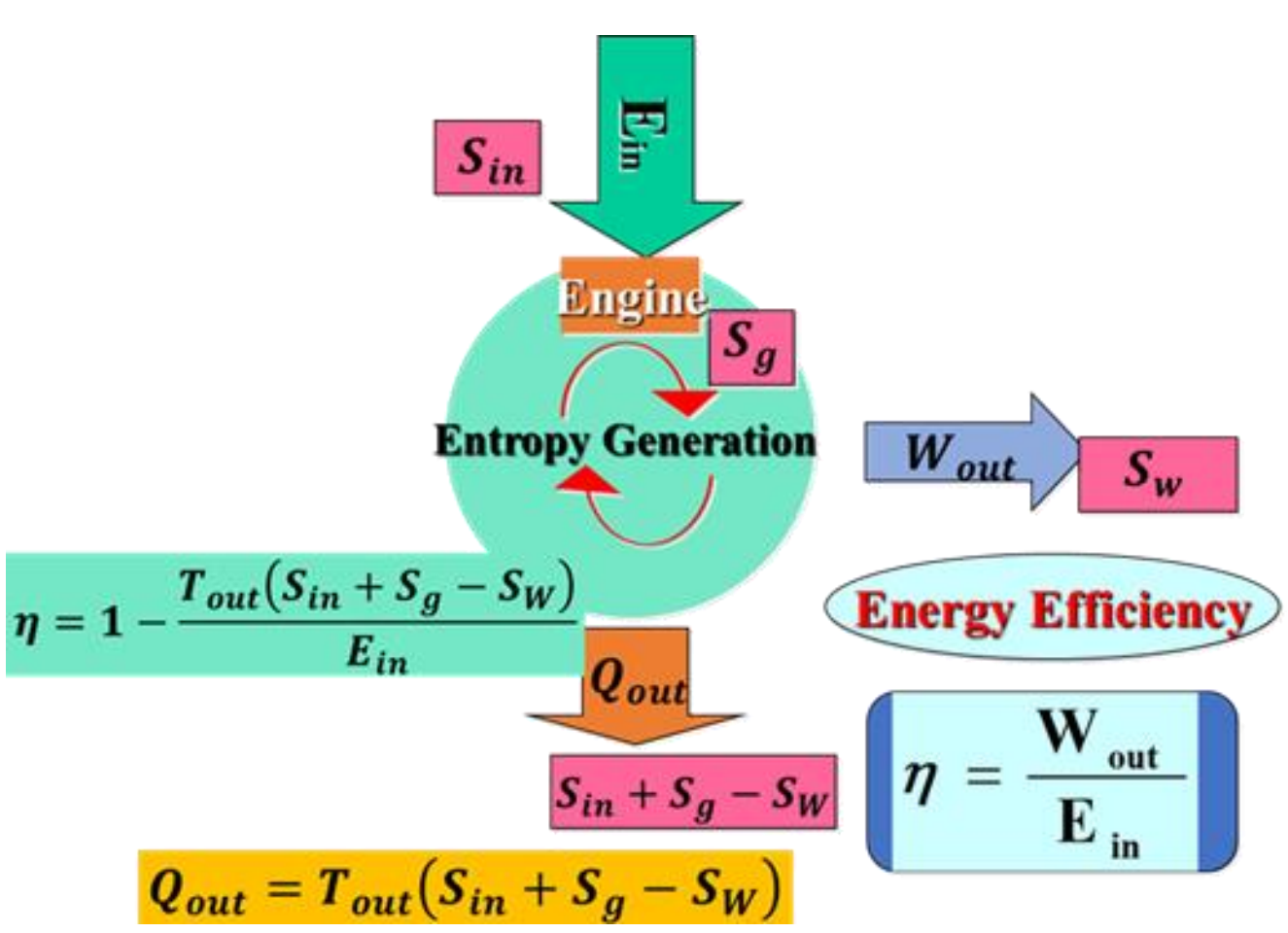

**Fig. 1. Energy and entropy flows entering, exiting, and generated by a powered system:**
This figure shows the flow diagram for a powered system. It forms the basis of the analysis in this study. The energy source $E_{in}$ is analyzed as the radiation energy $E_{in}^{\gamma}$ and the entropy discarding is only done by the waste heat $Q_{out}$.

According to the second law of thermodynamics, which implies that the total entropy can never decrease, i.e. $S_g \geq 0$ is satisfied. Furthermore, according to the first law of thermodynamics, which implies that the total energy must be conserved, $W_{out} = E_{in} - Q_{out}$ is satisfied, and by the definition of energy efficiency $\eta = W_{out}/E_{in}$, we have

$$\eta = \frac{E_{in} - Q_{out}}{E_{in}} = 1 - \frac{Q_{out}}{E_{in}}. \tag{3.3}$$

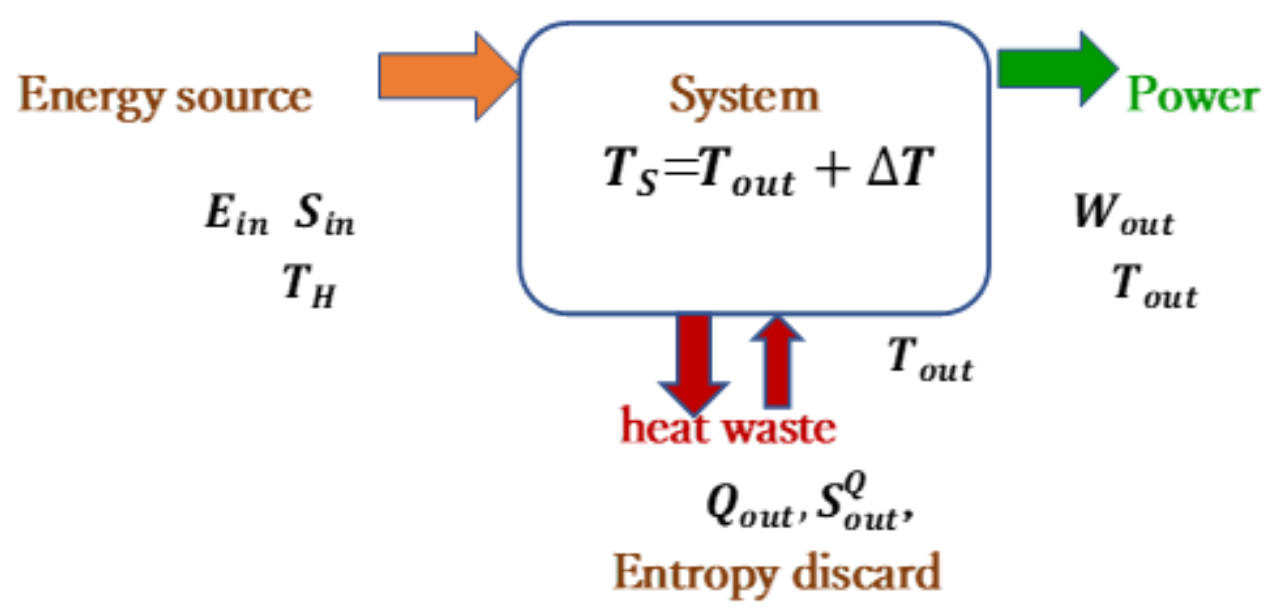

**Fig. 2. Schematic representation of the energy and entropy fluxes into and out of the system:** This diagram corresponds to the analysis in this paper for a *light-powered system* (such as photosynthesis), where the energy source is radiation (e.g., solar radiation) and entropy is discarded via waste heat. (In this figure, the temperatures of the inflow and outflow radiation of the system are represented by $T_{in}$ and $T_{out}$, $T_{out} + \Delta T_{out}$, respectively).

In the present study, it is assumed that in the process of entropy discard, the temperature difference $\Delta T$ between the system $\langle S \rangle$ and its external environment $\langle E \rangle$ is so small $(\Delta T \ll 1)$ that the net waste heat flow for entropy discard, expressed as the subtraction of the outward flow from the system to the environment and its reverse inward flow can be formulated by a first-order evaluation of $\Delta T/T$, which means the quasi-equilibrium condition between the system and the environment as a sink for entropy (Fig.2). As a result, the net outflow of entropy, $S^Q_{out}$, from the system is represented by the following Clausius formula.

$$\begin{aligned} S^Q_{out} &= C_V \ln(T_{out} + \Delta T) - C_V \ln T_{out}, \\ &= C_V \ln(1 + \Delta T/T_{out}) \cong C_V \, \Delta T/T_{out} = Q_{out}/T_{out}, \end{aligned} \tag{3.4}$$

where $C_V$ is the heat capacity at constant volume. Using Eqs. (3.1), (3.3), and (3.4), the following equation can be obtained:

$$\eta = 1 - \frac{T_{out}(S_{in}+S_g-S_W)}{E_{in}}. \quad (3.5)$$

From Eq. (3.5) and $S_g \geq 0$, the inequality $\eta \leq 1 - \frac{T_{out}(S_{in}-S_W)}{E_{in}}$ (the equality holds when$S_g = 0$) is evident, so the theoretical maximum energy efficiency $\eta_{max}$ corresponding to the case of an ideal reversible engine, wherein entropy is never generated, that is, $S_g = 0$ is given by

$$\eta_{max} = 1 - \frac{T_{out}(S_{in}-S_W)}{E_{in}}. \quad (3.6)$$

In the case of a general powered system, such as a heat engine, $S_W$ is almost zero in Eq. (3.6) , and $\eta_{max}$ is reduced to

$$\eta_{max} = 1 - \frac{T_{out}S_{in}}{E_{in}}. \quad (3.7)$$

When a heat engine is under the Carnot condition, i.e. the quasi-equilibrium condition between the heat source and the system which allows the first-order evaluation of $\Delta T/T_{in}$, then the Clausius formula $S_{out}^{Q} = Q_{in}/T_{in}$ holds (where $T_{in}$ is the energy source temperature) and its theoretical maximum energy efficiency $\eta_{max}$ given by Eq. (3.7) is obtained as follows, considering that the source energy $E_{in}$ is the heat energy $Q_{in}$,

$$\eta_{max} = 1 - \frac{T_{out}\frac{Q_{in}}{T_{in}}}{Q_{in}} = 1 - \frac{T_{out}}{T_{in}}, \quad (3.8)$$

which is Carnot efficiency.

In this paper, "theoretical maximum energy efficiency" and "ideal efficiency" are used interchangeably.

## 4. First-order evaluation of the ideal efficiency of a light-powered system at any wavelength, solid angle, and degree of polarization

The mechanism by which a light-powered system such as photosynthesis (an ensemble of molecules) absorbs energy from radiation (an ensemble of photons) and transfers entropy from a population of photons into the system is fundamentally different from the usual mechanism of entropy transfer accompanying the energy transfer between matter (an ensemble of molecules). In the latter case, the transfers are achieved macroscopically as heat transfer, by the collision of molecules or atoms at the microscopic level. In contrast, in the former case, they are achieved by absorption of a photon as a particle, by an electron. In the following discussion, the entropy change $\Delta S$ is analyzed in terms of the first-order evaluation by the number of photons (equilibrium approximation), so that $S_{in}(\lambda)$ is represented as the first-order evaluations by $\Delta f = \Delta N/G$.

In this section, the energy and entropy flow analysis performed in Section 3 is applied to a light-powered system that extracts power from radiation energy, and its ideal efficiency (maximum energy efficiency) is analyzed and formulated under various conditions. In the rest of this section, the quantities that can be obtained as a first-order evaluation of the photon numbers are denoted by a subscript (1) in the upper right-hand corner (e.g. $S_{in}^{\gamma(1)}$, $\Delta S^{\gamma(1)}$).

### 4.1. Basic formula

We first consider the case where the light is unpolarized and its light intensity (photon flux density) is independent of direction. That is, the number of photons absorbed by the light source system is assumed to be uniformly distributed within a certain solid angle_$\Omega$. Thus, Eq. (2.2) on $S(\lambda,\Omega)$ is a basis for the following entropy analysis. When $N_{in}^{\gamma}(\lambda)$ is the number of photons absorbed per second by a light-powered system from monochromatic light (a photon ensemble), the energy flow $E_{in}^{\gamma}(\lambda)$ and the entropy flow $S_{in}^{\gamma}(\lambda,\Omega)$ imported into the system per second can be expressed within the first-order evaluation by the number of photons as

$$E_{in}^{\gamma}(\lambda) = \frac{hc}{\lambda} N_{in}^{\gamma}(\lambda)\,, \tag{4.1}$$

$$\begin{aligned}
S_{in}^{\gamma(1)}(\lambda) &= -\Delta S^{\gamma(1)}\left(N_{in}^{\gamma}(\lambda),\lambda,\Omega\right) \\
&= \frac{\partial S^{\gamma}\left(f(\lambda,\Omega),G(\lambda,\Omega)\right)}{\partial N^{\gamma}(\lambda)}\left(-\Delta N^{\gamma}(\lambda)\right) \\
&= \frac{\partial S^{\gamma}\left(f(\lambda,\Omega),G(\lambda,\Omega)\right)}{\partial f(\lambda,\Omega)} \frac{\partial f(\lambda,\Omega)}{\partial N^{\gamma}(\lambda)}\left(-\Delta N^{\gamma}(\lambda)\right) \\
&= \frac{\partial S^{\gamma}(\lambda,\Omega)}{\partial f(\lambda,\Omega)} \frac{1}{G(\lambda,\Omega)} N_{in}^{\gamma}(\lambda) \\
&= k_B \ln\left(1+\frac{1}{f(\lambda,\Omega)}\right) N_{in}^{\gamma}(\lambda).
\end{aligned} \tag{4.2}$$

In Eq. (4.2), $\Delta S^{(1)}(\lambda,\Omega)$ represents the first-order evaluation of the change in entropy due to the change in number of photons $\Delta N^{\gamma}(\lambda)$ in an ensemble, then $S_{in}^{(1)}(\lambda,\Omega) = -\Delta S^{(1)}(\lambda,\Omega)$ and $N_{in}^{\gamma}(\lambda) = -\Delta N^{\gamma}(\lambda)$ hold, and it was used in the last equality Eq. (2.2).

### 4.2. Formula in the case of blackbody radiation

When the radiation (photon population) is in thermal equilibrium, i.e. blackbody radiation, at temperature $T_{in}$, $f(\lambda,\Omega=4\pi)$ is reduced to $f_{BB}(\lambda,T_{in}) = \frac{1}{\exp\left(\frac{hc}{\lambda k_B T_{in}}\right)-1}$, and thus $S_{in}^{\gamma}(\lambda,\Omega)$ is reduced to $S_{in}^{\gamma(1)}(\lambda,\Omega=4\pi) = \frac{hc}{\lambda T_{in}} N_{in}^{\gamma}(\lambda)$ from Eq. (4.2), and $E_{in}^{\gamma}(\lambda) = \frac{hc}{\lambda} N_{in}^{\gamma}(\lambda)$. Assuming that $S_W = 0$ in Eq.(3.6), the ideal efficiency $\eta_{max}^{\gamma}(\lambda,\Omega=4\pi)$ of the light-powered system becomes

$$\eta_{max,BB}^{\gamma}(\lambda,T_{in}) = 1 - \frac{T_{out}\frac{hc}{\lambda T_{in}} N_{in}^{\gamma}(\lambda)}{\frac{hc}{\lambda} N_{in}^{\gamma}(\lambda)} = 1 - \frac{T_{out}}{T_{in}}\,, \tag{4.3}$$

which is exactly the same as the formula for Carnot efficiency in a heat-powered system.

### 4.3. Formula of ideal efficiency for a light-powered system in general cases

General cases under non-equilibrium conditions, but with unpolarized light that is uniformly distributed in a given solid angle $\Omega$, can be analyzed using the proposed formulation as follows. By substituting Eq. (2.3) into Eq. (4.2), the entropy flow imported from a photon ensemble into a light-powered system is obtained as

$$S_{in}^{\gamma(1)}(\lambda, \Omega, \rho_N(\lambda)) = k_B \ln\left(1 + \frac{2\Omega\Delta\lambda}{\rho_N(\lambda)\, \lambda^4}\right) N_{in}^{\gamma}(\lambda). \tag{4.4}$$

Since the number density of a photon population $\rho_N(\lambda)$ in Eq. (4.4) is related to the photon flux density $n_\gamma(\lambda)$ by the formula $n_\gamma(\lambda) = (1/4)c\rho_N(\lambda)$, and using the photon flux density per wavelength $n_\gamma(\lambda)/\Delta\lambda$=$\widehat{n_\gamma}(\lambda)$, this formula is represented as

$$S_{in}^{\gamma(1)}\left(\lambda, \Omega, \widehat{n_\gamma}(\lambda)\right) = k_B \ln\left(1 + \frac{\Omega c}{2\lambda^4 \widehat{n_\gamma}(\lambda)}\right) N_{in}^{\gamma}(\lambda). \tag{4.5}$$

Substituting Eqs. (4.1) and (4.5) into Eq. (3.6) and assuming $S_W = 0$, the formula of ideal efficiency for light-powered system $\eta_{max}$ can be expressed as

$$\eta_{max}^{\gamma}\left(\lambda, \Omega, \widehat{n_\gamma}(\lambda)\right) = 1 - \frac{T_{out} S_{in}^{\gamma(1)}\left(\lambda,\Omega,\widehat{n_\gamma}(\lambda)\right)}{E_{in}^{\gamma}(\lambda)} = 1 - \frac{T_{out} k_B \ln\left(1 + \frac{\Omega c}{2\lambda^4 \widehat{n_\gamma}(\lambda)}\right)}{hc/\lambda}. \tag{4.6}$$

#### 4.3.1. Pseudo-radiation temperature $T_\gamma(\lambda, \Omega, \hat{n}(\lambda))$

The term extracted in Eq. (4.6) with temperature dimension is expressed as

$$T_\gamma(\lambda, \Omega, \hat{n}(\lambda)) = \frac{E_{in}^{\gamma}(\lambda)}{S_{in}^{\gamma(1)}\left(\lambda,\Omega,\widehat{n_\gamma}(\lambda)\right)} = \frac{hc/\lambda}{k_B \ln\left(1 + \frac{\Omega c}{2\lambda^4 \widehat{n_\gamma}(\lambda)}\right)}, \tag{4.7}$$

and Eq. (4.6) is expressed as

$$\eta_{max}^{\gamma}\left(\lambda, \Omega, \widehat{n_\gamma}(\lambda)\right) = 1 - \frac{T_{out}}{T_\gamma(\lambda,\Omega,\hat{n}(\lambda))}. \tag{4.8}$$

It can be shown that when the solid angle range $\Omega$ in Eq. (4.7) is 4π (i.e., isotropic), $T_\gamma(\lambda, \hat{n}(\lambda), \Omega)$ coincides with the effective temperature $T_D(\lambda)$ of the radiation extracted from Planck's spectral distribution law, as explained in the next subsection 4.3.2. note below. Although $T_\gamma(\lambda, \hat{n}(\lambda), \Omega)$ and $T_D(\lambda)$=$T_\gamma(\lambda, \hat{n}(\lambda), 4\pi)$ have a temperature dimension, as the wavelength dependence indicates, they are not true temperatures defined at an equilibrium state, but rather quantities that should be called pseudo-temperatures. However, since the entropy change $\Delta S^{\gamma}(\lambda, \Omega)$ formulated in Eq. (4.2) is given by the first-order evaluation by the photon number change $\Delta N^{\gamma}(\lambda)$, the formula $1/T_\gamma(\lambda, \hat{n}(\lambda), \Omega) =$ $ds(\lambda,\Omega)/\,du(\lambda)$ holds, as in the case of equilibrium thermodynamics. Its details are shown in Appendix B.

### 4.3.2. Effective temperature $T_D(\lambda)$ in the conventional studies

In Section 2, the number flux per wavelength of blackbody radiation $\hat{n}_{BB}(\lambda,T)$ was obtained as follows,

$$\hat{n}_{BB}(\lambda,T)=\frac{2\pi c}{\lambda^4}\frac{1}{\exp\left(\frac{hc}{\lambda k_B T}\right)-1}. \tag{4.9}$$

Replacing $\hat{n}_{BB}(\lambda,T)$ by any $\hat{n}(\lambda)$ and solving Eq. (4.9) for $T$, the effective temperature $T=T_D$ is given by

$$T_D\big(\lambda,\hat{n}(\lambda)\big)=\frac{hc/\lambda}{k_B\ln\left(1+\frac{2\pi c}{\lambda^4\hat{n}(\lambda)}\right)}. \tag{4.10}$$

The usual procedure for analyzing the maximum energy efficiency $\eta_{max}\big(\lambda,\hat{n}(\lambda)\big)$ of photosynthesis using the effective temperature $T_D\big(\lambda,\hat{n}(\lambda)\big)$ given by Eq. (4.10) has often been criticized because $\eta_{max}^{\gamma}(\lambda)$ is calculated by substituting $T_D(\lambda)$ into the Carnot efficiency formula by treating $T_D\big(\lambda,\hat{n}(\lambda)\big)$ as a temperature like a standard thermal equilibrium temperature. In contrast, in the formulation proposed in this study, $\eta_{max}^{\gamma}(\lambda)$ is consistently calculated in terms of energy and entropy analysis, and subsequently, $T_{\gamma}(\lambda)$ is extracted as a quantity having a temperature dimension as Eq. (4.7). Thus, $T_{\gamma}(\lambda)$ need not be the true temperature defined only by an equilibrium condition. The essential reason for the agreement between Eq. (4.7) obtained by the theoretical method in this study and Eq. (4.10) obtained by the conventional method is discussed in Section 8.3.

The radiation temperature formula defined including solid angle in the book Statistical Physics (Second Revised and Enlarged Edition (1969)) written by L.D. Landau and E.M. Lifshitz) [24] can be found to be essentially the same as Eq. (4.7). (A detailed explanation is given in Appendix D.) However, this temperature is obtained as an extension of the temperature of blackbody radiation, as these authors themselves state. In previous studies, the ideal efficiency of a light-powered system such as photosynthesis has only been obtained by automatically applying this temperature to the Carnot efficiency formula without any reason. We have no evidence why this pseudo-temperature can be automatically applied to the Carnot efficiency formula to obtain the ideal efficiency of a light-powered system.

### 4.3.3. Lower bound of the photon flux density for extracting power

Solving Eq. (4.7) for $\hat{n}(\lambda)$, the photon flux density per wavelength, whether in equilibrium or non-equilibrium conditions, gives the following formula:

$$\widehat{n_\gamma}(\lambda) = \frac{\Omega c}{2\lambda^4}\frac{1}{\exp\left(\frac{hc}{\lambda k_B T_\gamma}\right)-1}. \tag{4.11}$$

Eq. (4.11) implies the existence of a threshold value of the photon flux density $\hat{n}_\gamma(\lambda)$ for extracting power; i.e., there is a lower bound $\hat{n}_0(\lambda)$. From Eq. (4.8), it is clear that unless $T_\gamma(\lambda, \hat{n}(\lambda)) \geq T_{out}$ is satisfied, $\eta^\gamma_{max}(\lambda, \Omega, \hat{n}(\lambda))$ becomes negative and no power can be extracted. By solving the inequality $T_\gamma(\lambda, \hat{n}(\lambda)) \geq T_{out}$ for the photon flux density $\widehat{n_0}(\lambda)$, t we obtain:

$$\widehat{n_\gamma}(\lambda) \geq \frac{\Omega c}{2\lambda^4}\frac{1}{\exp\left(\frac{hc}{\lambda k_B T_\gamma}\right)} = \widehat{n_0}(\lambda). \tag{4.12}$$

More than half a century ago, Duysens showed that the efficiency of energy-converting devices approaches zero as the intensity of light (photon flux density) approaches zero, due to the second law of thermodynamics [25]. After being exposed to criticisms from researchers such as Kahn [26], Duysens refuted Kahn's argument [27]. Now that Eq. (4.12) is found to hold as above, it is obvious that Duysens' result is correct. Duysens also mentioned that the calculated efficiency is achieved only in a reversible process, while the actual efficiency of a non-reversible process is lower than this calculated efficiency [27].

**4.3.4. Polarized light**

This section presents a mathematical formula for the theoretical maximum energy efficiency of a light-powered system whose energy source is monochromatic light with wavelength λ, arbitrary solid angle Ω, and polarization parameter $\alpha_i (i = 1,2)$. When $N^\gamma_{in}(\lambda)$ photons are absorbed by a light-powered system from monochromatic light (a photon population), the energy $E^\gamma_{in}(\lambda)$ and entropy $S^\gamma_{in}(\lambda, \Omega, \alpha_1, \alpha_2)$ imported into the system can be expressed as

$$E^\gamma_{in}(\lambda) = \frac{hc}{\lambda} N^\gamma_{in}(\lambda) , \tag{4.13}$$

$$\begin{aligned} S^{\gamma(1)}_{in}(\lambda, \Omega, \alpha_1, \alpha_2) &= -\Delta S^{(1)}(\lambda, \Omega, \alpha_1, \alpha_2) \\ &= \left\{\frac{\partial}{\partial N^\gamma(\lambda)} \sum_{i=1}^{2} S(\lambda, \Omega, \alpha_i)\right\}\left(-\Delta N^\gamma(\lambda)\right) \\ &= k_B \sum_{i=1}^{2} \alpha_i \ln\left(1 + \frac{1}{2\alpha_i f(\lambda, \Omega)}\right) N^\gamma_{in}(\lambda). \end{aligned} \tag{4.14}$$

The expansion in lines 2–3 in Eq. (4.14) used Eq. (4.2).

Eq. (4.14) gives the following formulae.

(1) For unpolarized light ($\alpha_1 = \alpha_2 = 1/2$),

$$S^{\gamma(1)}_{in}(\lambda, \Omega) = k_B \ln\left(1 + \frac{1}{f(\lambda, \Omega)}\right) N^\gamma_{in}(\lambda), \tag{4.15}$$

which coincides with Eq. (4.2).

(2) For completely polarized light ($\alpha_1 = 1, \alpha_2 = 0$ or $\alpha_2 = 1, \alpha_1 = 0$),

$$S_{in}^{\gamma(1)}(\lambda,\Omega) = k_B \ln\left(1+\frac{1}{2f(\lambda,\Omega)}\right) N_{in}^{\gamma}(\lambda), \tag{4.16}$$

Eq. (4.16) coincides with that obtained by eliminating the numerical factor 2 as the spin degree of freedom of the photon from Eq. (4.2).

From Eqs. (3.7), (4.5) and (4.13), (4.14), the general formula for the ideal efficiency, including polarization, is finally obtained as

$$\eta_{max}^{\gamma}\left(\lambda,\Omega,\widehat{n_\gamma}(\lambda),\alpha_1,\alpha_2\right) = 1 - \frac{T_{out}\boldsymbol{S}_{in}^{\gamma(1)}(\lambda,\Omega,\widehat{n_\gamma}(\lambda),\alpha_1,\alpha_2)}{\boldsymbol{E}_{in}^{\gamma}(\lambda)}$$

$$= 1 - \frac{T_{out}k_B\sum_{i=1}^{2}\alpha_i \ln\left(1+\frac{\Omega c}{4\alpha_i\lambda^4\widehat{n_\gamma}(\lambda)}\right)}{hc/\lambda}, \tag{4.17}$$

and the radiation pseudo-temperature is generally given by

$$T_\gamma\left(\lambda,\Omega,\widehat{n_\gamma}(\lambda),\alpha_1,\alpha_2\right) = \frac{hc/\lambda}{k_B\sum_{i=1}^{2}\alpha_i \ln\left(1+\frac{\Omega c}{4\alpha_i\lambda^4\widehat{n_\gamma}(\lambda)}\right)} . \tag{4.18}$$

It is easy to see that Eq. (4.18) becomes Eq. (4.7) when $\alpha_i = 1/2\ (i = 1,2)$.

Section 5.3 of this paper shows, based on Eq.(4.10), how to use equation (5.1) to find the effective temperature of sunlight at the Earth's surface, $T_D(670\text{nm}) = 1490\text{K}$, and then substitute it into the Carnot efficiency to calculate the ideal efficiency, $\eta_{max}(670\text{nm}) = 0.799$. This method necessarily allows evaluation only for polarization 0 due to the method of deriving the effective temperature, but equation (4.18) derived in this study allows evaluation of the radiation temperature and ideal efficiency at any degree of polarization. In reality, sunlight is polarized at the Earth's surface due to atmospheric scattering, so equation (4.18) allows for a more general and realistic evaluation than the effective temperature method.

Figure 3 shows the dependence of the radiation temperature $T_\gamma(\alpha_1, \Omega = 4\pi, \lambda = 670nm)$ and the ideal efficiency $\eta_{max}^{\gamma}(\alpha_1, \Omega = 4\pi, \lambda = 670nm, T_{out} = 300\text{K})$ given by Eqs. (4.17) and (4.18) on the polarization parameter $\alpha_1$. In this figure, when $\alpha_1 = 0.5$, i.e., when the degree of polarization is 0, the values coincide with the calculated values based on the effective temperature. Looking at the graph of $\eta_{max}^{\gamma}$, we can see that this is almost the same as the approximately 1% difference in ideal efficiency between completely polarized ($\alpha_1 = 0,1$) and unpolarized light ($\alpha_1 = 1/2$), which was the only thing that could be analyzed in previous studies [21].

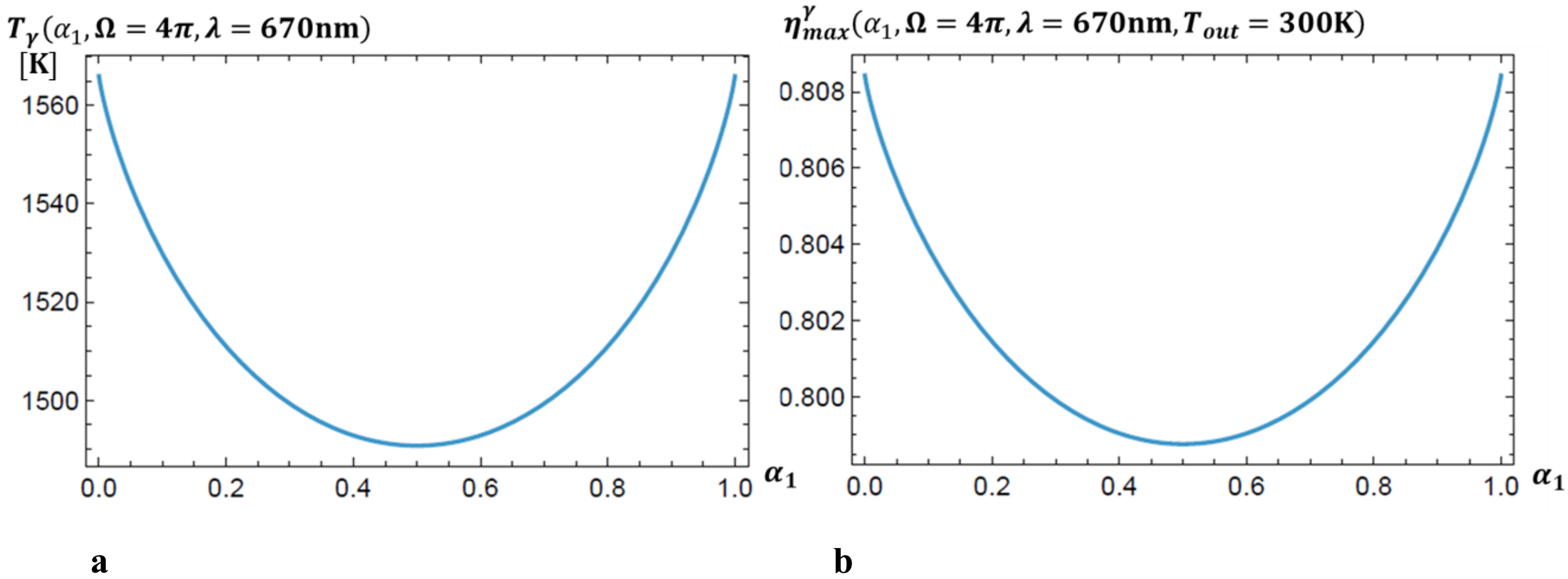

**Fig. 3. Dependence of (a) The radiation temperature $T_\gamma(\alpha_1, \Omega = 4\pi, \lambda = 670nm)$ and (b) The ideal efficiency of a *light-powered system* $\eta^\gamma_{max}(\alpha_1, \Omega = 4\pi, \lambda = 670nm, T_{out} = 300\text{K})$ on the polarization parameter $\alpha_1$.** When the polarization parameters $\alpha_1 = 0.5$ $(\alpha_2 = 1 - \alpha_1 = 0.5)$, the radiation temperature $T_\gamma(\alpha_1 = 0.5, 670\text{nm})$ and the ideal efficiency $\eta^\gamma_{max}(\alpha_1 = 0.5, 670\text{nm}, T_{out} = 300\text{ K})$ coincide with the effective temperature $T_D = 1490\text{K}$ obtained from Eq.(5.1) based on blackbody radiation and the ideal efficiency $\eta^\gamma_{max}(T_D, T_{out} = 300\text{ K}) = 0.799$ obtained using this temperature, respectively.

## 5. Ideal efficiency of terrestrial photosynthesis under diluted solar radiation

The entropy in the configuration space is increased due to the dilution of photons from the Sun to the Earth (Fig.4) [28]. As a result, the ideal efficiency (quasi-Carnot efficiency) would naturally be expected to decrease. To analyze the ideal efficiency of photosynthesis on Earth, the dilution rate of solar radiation from the Sun to the Earth must be taken into account. Conventional methods have analyzed the ideal efficiency by incorporating the dilution rate into the radiation temperature . In this method, the radiation temperature is taken from a spectrum deviated from blackbody radiation by dilution, which inevitably takes the form of a wavelength-dependent temperature (called the effective temperature). This has led to the suspicion that this is not the true temperature and the method based on such a pseudo-temperature may be incorrect. In contrast, in this study, the ideal efficiency analysis was performed by incorporating the dilution rate into the energy and entropy flow analysis in Section 3 in a consistent manner, rather than taking out the quantity of temperature from the beginning. This section explains how each of these two methods derives the ideal efficiency of the terrestrial photosynthetic system under diluted solar radiation.

### 5.1. Derivation using the effective temperature

In the study of the ideal efficiency for photosynthesis on Earth, the effective temperature of the diluted solar radiation for a given wavelength using the dilution rate of the photon flux density per wavelength $\hat{n}(\lambda)$ was originally derived by Duysens [5] and later sorted out by P.T. Landsberg and G. Tonge [29]. To summarize this process compactly, when $\hat{n}_D(\lambda) = d\hat{n}(\lambda)$ ($d$ is the dilution rate) is given as the dilution effect, the effective temperature $T_D^{\gamma}(\lambda)$ is defined by the following procedure:

$$\hat{n}_D(\lambda) = \frac{2\pi c}{\lambda^4}\frac{1}{\exp\left(\frac{hc}{\lambda k_B T_D^{\gamma}}\right)-1} = d\hat{n}(\lambda) = d\frac{2\pi c}{\lambda^4}\frac{1}{\exp\left(\frac{hc}{\lambda k_B T_{sun}}\right)-1} \quad . \tag{5.1}$$

Solving Eq. (5.1) for $T_D^{\gamma}(\lambda)$, we obtain

$$T_D^{\gamma}(\lambda) = \frac{hc/\lambda k_B}{\ln\left\{\frac{1}{d}\left\{\exp\left(\frac{hc}{\lambda k_B T_{sun}}\right)-1\right\}+1\right\}}. \tag{5.2}$$

Substituting $T_{in} = T_D^{\gamma}(\lambda)$ into the Carnot formula $\eta_C = 1 - T_{out}/T_{in}$(where $T_{in}$ and $T_{out}$ are the temperature of an energy source and a waste energy sink, respectively), and applying the diluted solar radiation on the Earth $d = R^2/D^2$ (where $R$ is the Sun's radius and $D$ is the Sun-Earth distance), we obtain the theoretical maximum energy efficiency $\eta_{max}$ of terrestrial photosynthesis as follows:

$$\eta_{max}^{\gamma D}(\lambda) = 1 - \frac{T_{out}}{T_D^{\gamma}(\lambda)} = 1 - \frac{T_{out}}{hc/\lambda} k_B \ln\left\{\frac{D^2}{R^2}\left\{\exp\left(\frac{hc}{\lambda k_B T_{sun}}\right) - 1\right\} + 1\right\}. \tag{5.3}$$

Using $\lambda = 670$ nm as an example, together with $R = 6.96 \times 10^8$ m, $D = 1.50 \times 10^{11}$ m, $T_S =$ 5800 K, $T_S = 300$ K, $h = 6.63 \times 10^{-34}$Js, c = $3.00 \times 10^8$ m/s, and $k_B = 1.38 \times 10^{-23}$ J/K, the values of $T_D(\lambda)$ and $\eta_{max}^{\gamma D}(\lambda = 670\text{ nm})$ are obtained as 1490 K and 0.799, respectively.

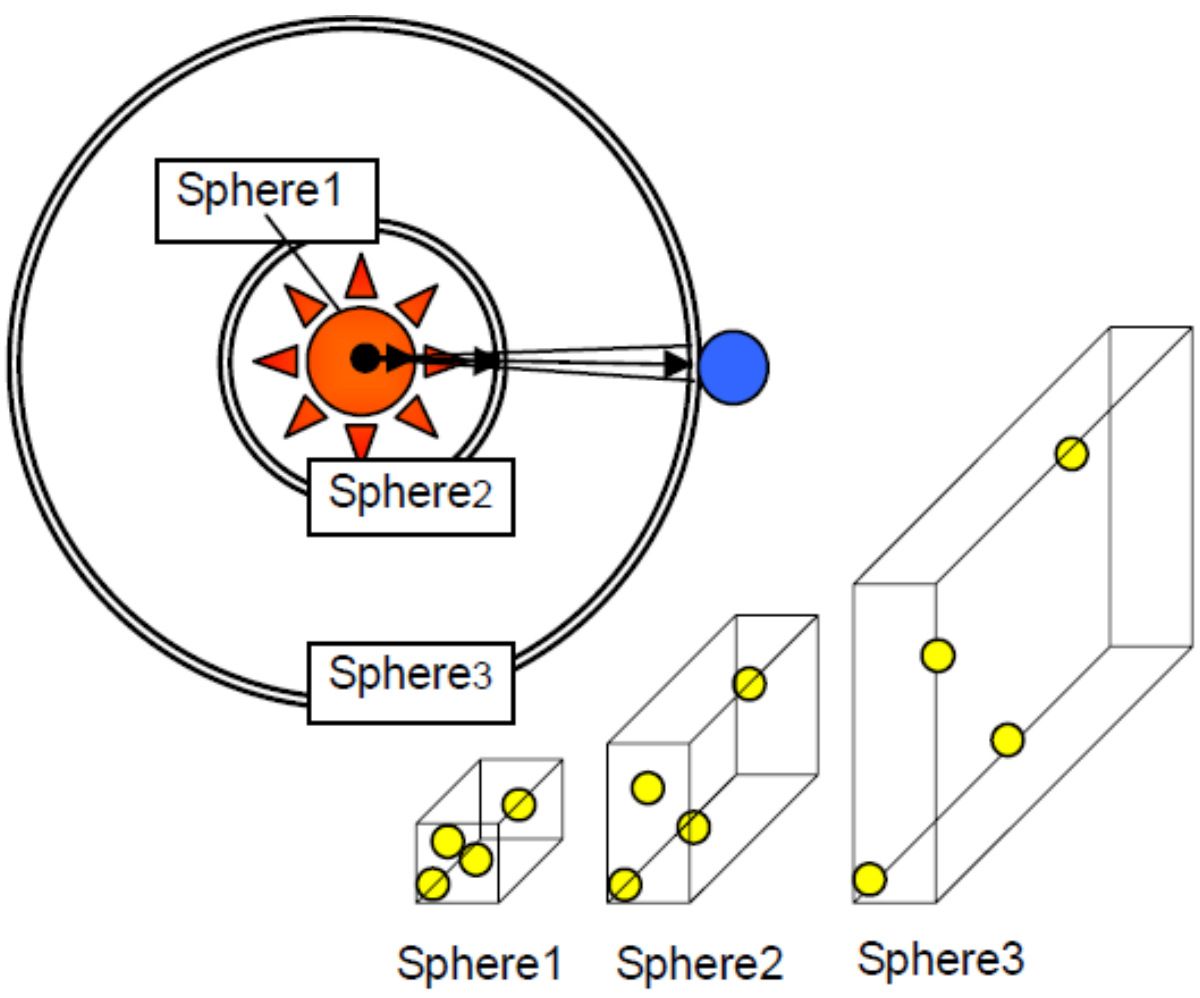

**Fig. 4. Increased entropy in the configuring space due to photon dilution:** This picture is depicted for intuitive understanding of the increased entropy in configuring space due to photon dilution from the Sun to the Earth [28]. This picture is taken from Toshifumi Konno's 2003 graduate thesis.

### 5.2. Derivation based on energy and entropy flow analysis

Terrestrial plants use the diluted solar radiation transported from the Sun for photosynthesis. One question is whether the entropy of solar radiation (photon population) increases before entering the Earth's atmosphere. The answer is no. According to Liouville's theorem in the absence of interaction, the configuration-space entropy of solar energy at the Earth increases with dilution, while the momentum-space entropy decreases with parallel orientation (specifically, a decrease in the solid angle Ω), and these quantities cancel each other out completely. The essence of this understanding is that the photons of radiation from the Sun to the outermost layers of the Earth's atmosphere interact with almost nothing as they pass through the near vacuum of space before entering the Earth's atmosphere. The fact that the increase in the configuration-space volume is cancelled out by the decrease in the momentum-space volume and that the phase-space volume remains invariant can be understood, to an approximation using Eq. (2.3) as follows; the spherical shell volume $V$ becomes $D^2/R^2$ times larger and the solid angle $\Omega$ becomes $\Omega_{earth}/\Omega_{sun} = (\pi R^2/D^2)/(2\pi) = (1/2)R^2/D^2$ times smaller, so we can see that the two almost cancel out, but according to this understanding, the total volume of phase space is reduced, albeit by only 1/2. A rigorous analysis of this cancellation is given by the present author using elementary geometry in Appendix C.

After entering the Earth's atmosphere and undergoing random scattering within the atmosphere and clouds to broaden the direction of momentum, an increase in the configuration entropy of the photons of solar radiation due to their dilution becomes apparent. Consequently, the ideal efficiency of the work extracted from the solar radiation is reduced. In this section, the entropy of the solar radiation after entering the Earth's atmosphere and the theoretical maximum energy efficiency $\eta_{max}^{D}$ are analyzed, considering that it interacts randomly with the atmospheric molecules so that its directions are scattered throughout the solid angle of 4π. In this analysis, these scatterings are assumed to be independent of wavelength $\lambda$ and uniform, independent of polarization and solid angle. I In reality, solar radiation is scattered non-uniformly in the atmosphere with a dependence on the wavelength $\lambda$ (Rayleigh scattering, etc.) and results in a certain polarization and solid angle dependence. A realistic analysis that takes these conditions into account, which we are currently working on, is described in the last section.

The dilution effect of the photon ensemble coming from the Sun manifested by these scattering processes, is represented by the dilution ratio $R^2/D^2$, and thus the photon flux density $\hat{n}_D(\lambda)$ becomes

$$\hat{n}_D(\lambda, T_{sun}) = \frac{R^2}{D^2}\hat{n}_{BB}(\lambda, T_{sun}). \tag{5.4}$$

Substituting $\hat{n}_{BB}(\lambda, T_S)$ from (4.9), obtained by Planck's law, into Eq. (5.4), the following equation is obtained:

$$\hat{n}_D(\lambda) = \frac{R^2}{D^2}\,\frac{2\pi c}{\lambda^4}\frac{1}{\exp\left(\frac{hc}{\lambda k_B T_{sun}}\right)-1}. \tag{5.5}$$

Substituting $\hat{n}_D(\lambda)$ for $\hat{n}_\gamma(\lambda)$ in Eq. (4.5) with the solid angle $\Omega = 4\pi$, the following formula is obtained:

$$S_{in}^{\gamma(1)}\left(\lambda, \Omega = 4\pi, \widehat{n_D}(\lambda)\right) = k_B \ln\left(1 + \frac{2\pi c}{\lambda^4 \widehat{n_D}(\lambda)}\right) N_{in}^{\gamma}(\lambda)$$

$$= k_B \ln\left\{1 + \frac{D^2}{R^2}\left(\frac{hc}{\lambda k_B T_{sun}} - 1\right)\right\} N_{in}^{\gamma}(\lambda). \tag{5.6}$$

Using Eq. (4.6), the following formula is obtained as the ideal efficiency:

$$\eta_{max}^{\gamma}\left(\lambda, \Omega = 4\pi, \widehat{n_D}(\lambda)\right) = 1 - \frac{T_{out} S_{in}^{(1)}\left(\lambda, \Omega = 4\pi, \widehat{n_D}(\lambda)\right)}{E_{in}(\lambda)}$$

$$= 1 - \frac{T_{out}}{hc/\lambda} k_B \ln\left\{1 + \frac{D^2}{R^2}\left(e^{hc/\lambda k_B T_{sun}} - 1\right)\right\}. \tag{5.7}$$

Eq. (5.7), based on the first principle of thermodynamics, coincides with Eq. (5.3) obtained using the effective temperature $T_D^{\gamma}(\lambda)$ and the Carnot efficiency formula. Both derivation methods assume a reversible process, but since the latter substitutes the temperature of the energy source (radiation) in the Carnot efficiency equation, it can be said that a state of thermal equilibrium, the condition defining the temperature, is also assumed with respect to the energy source. In other words, Carnot efficiency is based on two assumptions: the reversibility condition and the thermal equilibrium condition for the energy source, the latter of which gives the temperature of the energy source. However, it should be noted that diluted terrestrial solar radiation is not in thermal equilibrium, so the Carnot efficiency formula cannot be used a priori. In this respect, the analysis using the effective temperature $T_D^{\gamma}(\lambda)$has been critically discussed because it automatically assigns a non-equilibrium temperature (i.e., not the true temperature) to the Carnot efficiency formula without any basis. This is due to the fact that the order of the analysis is the opposite of the correct order. In contrast, the method used in this study was to first derive a general expression for the theoretical maximum energy efficiency $\eta_{max}$, and the Boltzmann-type factor giving the concentration ratio of excited-to-ground state pigment-molecules by energy and entropy flow analysis based on the second law of thermodynamics, and then to extract the temperature dimensional quantity from the formula of $\eta_{max}$ given by Eq. (4.6) as Eq. (4.7).

According to Eq. (4.7), $T_\gamma\left(\lambda, \hat{n}_\gamma(\lambda), \Omega\right)$ corresponding to the diluted photon flux density $\hat{n}_D(\lambda)$ can be expressed as

$$T_\gamma(\lambda, \hat{n}_D(\lambda), \Omega = 4\pi) = \frac{hc/\lambda}{k_B \ln\left(1+\frac{8\pi c}{\lambda^4 \hat{n}_D(\lambda)}\right)} = \frac{hc/\lambda}{k_B \ln\left\{1+\frac{D^2}{R^2}\left(e^{hc/\lambda k_B T_{sun}} - 1\right)\right\}}, \tag{5.8}$$

which is equal to the effective temperature $T_D(\lambda)$ given by Eq. (5.2) as $d = R^2/D^2$.
When $D = R$ on the solar surface, there is no dilution, Eq. (5.8) yields $T_\gamma\left(\lambda, \hat{n}_D(\lambda, T_S)\right) = T_S$, and Eq. (5.7) yields $\eta_{max}\left(\lambda, \Omega = 4\pi, \widehat{n_D}(\lambda, T_S)\right) = 1 - \frac{T_{out}}{T_S}$, which are expected results.

From Eqs. (5.7) and (5.8) we obtain the following formula, which is similar to the Carnot efficiency formula:

$$\eta_{max}^{\gamma}(\lambda, , \widehat{n_D}(\lambda), \Omega = 4\pi) = 1 - \frac{T_{out}}{T_\gamma(\lambda, \hat{n}_D(\lambda), \Omega=4\pi)}. \tag{5.9}$$

Although Eq. (5.9) is the same as the formula $\eta_{max}^{\gamma D}(\lambda) = 1 - T_{out}/T_D(\lambda)$ in Eq. (5.3) obtained by substituting the effective temperature $T_D^{\gamma}(\lambda)$ into the Carnot efficiency formula, their logics are essentially different from each other as mentioned above.
Although a certain experiment with artificial light sources claimed, by applying a radiation temperature based only on photon flux density (i.e. light intensity), that the experimental result of energy efficiency $\eta$ exceeds the ideal efficiency, and thermodynamic principles might not be applicable to photonic systems such as photosynthetic photosystems [14], such a misunderstanding may have arisen from overlooking the condition of artificial sources, whose solid angle Ω is typically much smaller than 4π. As a result, their total entropy is smaller and $T_\gamma(\lambda, \hat{n}_D(\lambda, T_S), \Omega < 4\pi)$ obtained by Eq. (4.7) is higher. If its solid angle Ω had been correctly considered, it would have been understood that the radiation temperature $T_\gamma(\lambda, \hat{n}_D(\lambda, T_S), \Omega)$ formulated in this study was correctly estimated and thermodynamic principles such as the second law of thermodynamics were not violated. A quantitative analysis that the experimentally obtained high efficiency does not contradict the second law of thermodynamics by considering the solid angle of the laser light is given in the next section 5.3 using Eq. (4.7).
The above analysis is based only on the dilution of solar radiation as it travels from the Sun to the Earth, and does not take into account more real conditions for terrestrial solar radiation. In reality, the sunlight that reaches the Earth is absorbed and reflected in the atmosphere before it reaches the Earth's surface, further reducing the solar intensity (photon flux density) and increasing the entropy of the radiation, thus reducing the energy efficiency of the terrestrial light-powered system that uses it.

### 5.3. Examples of actual numerical values obtained by two derivation methods

The method described in 5.1 derives the ideal efficiency by applying the effective temperature$T_D(\lambda)$=$T_\gamma(\lambda, \hat{n}(\lambda), \Omega = 4\pi)$, obtained by fitting the light intensity (photon number flux) of monochromatic light to the Planck radiation spectrum under equilibrium conditions, to the Carnot efficiency, while the method used in this study described in 5.2 is a general derivation formulated as Eq. (4.18), using energy and entropy flow analysis of monochromatic light, which is not limited to equilibrium conditions and includes not only light intensity but also solid angle and degree of polarization, from which the general radiation temperature $T_\gamma(\lambda, \hat{n}(\lambda), \Omega, \alpha_i)$ is extracted from it.

An example of the comparison of the two derivation methods is the calculation of the energy efficiency $\eta$ of photosynthesis obtained from experiments using artificial light such as lasers. For example, in an experiment measuring the excitation energy transfer efficiency using LHCI (Light Harvesting Complex) purified from higher plants, laser light, and PSI (Photosystem I) with a photon flux density equivalent to a wavelength λ = 670 nm and an effective temperature of 1100 K, the thermodynamic efficiency was reported to be at least 0.96, which is significantly higher than the value of 0.73 of the ideal efficiency obtained by the effective temperature $T_\gamma(\lambda)$ described in 5.1, and it has been reported that the ideal efficiency of photosynthesis exceeds the Carnot efficiency of thermodynamics [14]. On the other hand, this experimental result can be understood without contradiction by the generalized radiation temperature described in 5.2, formulated as Eq. (4.18), as follows. In its analysis, the quantum coherence of photons from a laser source is not considered.

The following general formulation of the radiation temperature $T_\gamma$ including a solid angle $\Omega$ is given by Eq. (4.7)

$$T_\gamma(\lambda, \Omega, \hat{n}(\lambda)) = \frac{hc/\lambda}{k_B \ln\left(1+\frac{\Omega c}{2\lambda^4 \hat{n}_\gamma(\lambda)}\right)} \qquad (5.10)$$

After a little calculation, the relationship between two radiation temperatures given by the same photon flux density $\widehat{n_\gamma}(\lambda)$) but different solid angles $\Omega$ and $\Omega'$ can be obtained as follows.

$$T_\gamma\left(\lambda, \Omega', \widehat{n_\gamma}(\lambda)\right) = \frac{hc/\lambda k_B}{\ln\left[1+\frac{\Omega'}{\Omega}\left\{\exp\left(\frac{hc}{\lambda k_B T_\gamma\left(\lambda,\Omega,\widehat{n_\gamma}(\lambda)\right)}\right)-1\right\}\right]} \qquad (5.11)$$

When the radiation temperature at $\lambda = 670\text{nm}$, $\Omega = 4\pi$ (i.e., the effective temperature), is 1100[K], as set in the experiment [14], the radiation temperature of artificial light, such as laser light, with the same photon flux density $\widehat{n_\gamma}(\lambda)$ can be calculated using the necessary constant values. Here, based on the data of the current general divergence angle $\theta$ of the laser light, we calculate the radiation temperature for (i) $\theta_1 = 0.5\text{mrad}$ and (ii) $\theta_2 = 0.1\text{mrad}$ using the equation $\Omega = 2\pi(1 - \cos\theta) \sim \theta^2 \pi$ $(\theta \ll 1)$. In this case, $\Omega'/\Omega = \theta^2/4$ in Eq. (5.11), and the following calculations can be made.

(i) $\theta_1 = 0.5\text{mrad}$

$$T_\gamma\left(670nm, \Omega_1, \widehat{n_\gamma}(670nm)\right) = 7150[\mathrm{K}] \tag{5.12}$$

$$\eta_{max}\left(670nm, \Omega_1, \widehat{n_\gamma}(670nm)\right) = 1 - \frac{T_{out}}{T_\gamma\left(670nm, \Omega_1, \widehat{n_\gamma}(670nm)\right)}$$

$$= 1 - \frac{300[\mathrm{K}]}{7150[\mathrm{K}]}$$

$$= 0.958 \ \ (95.8\%) \tag{5.13}$$

(ii) $\theta_2 = 0.1\mathrm{mrad}$

$$T_\gamma\left(670nm, \Omega_2, \widehat{n_\gamma}(670nm)\right) = 37700[\mathrm{K}] \tag{5.14}$$

$$\eta_{max}\left(670nm, \Omega_2, \widehat{n_\gamma}(670nm)\right) = 1 - \frac{T_{out}}{T_\gamma\left(670nm, \Omega_2, \widehat{n_\gamma}(670nm)\right)}$$

$$= 1 - \frac{300[\mathrm{K}]}{37700[\mathrm{K}]}$$

$$= 0.992 \ (99.2\%) \tag{5.15}$$

From the above analysis, which also takes into account the solid angle $\Omega$ of the artificial lighting used, it is found that the experimental results can be understood without contradiction.

In this example, because the conventional ideal efficiency formulation described in Section 5.1, which used the effective temperature derived from blackbody radiation (equilibrium radiation), implicitly assumed a solid angle of $\Omega = 4\pi$, it was unable to correctly calculate the ideal efficiency for laser light with a much smaller solid angle $\Omega$. In contrast, we found that the formulation described in Section 5.2, which is not limited to equilibrium states and is based on the energy-entropy flow analysis of monochromatic light, which considers not only light intensity but also solid angle and polarization angle, gives us a correct ideal efficiency that does not contradict experimental values. Figure 5. shows the graphs of radiation temperature $T_\gamma$and ideal efficiency $\eta_{max}^{\gamma}$ as a function of the polarization parameter $\alpha_1$ using Eq. (4.18) formulated in this study for lights at three different solid angles of $\Omega = 4\pi[\mathrm{sr}]$, $\Omega = 2.5 \times 10^{-7}\pi[\mathrm{sr}]$, and $\Omega = 1.0 \times 10^{-8}\pi[\mathrm{sr}]$, respectively, at a wavelength of 670 nm. These results show that the dependence of the ideal efficiency on the degree of polarization is smaller than its dependence on the solid angle, and show that experiments with laser light at small solid angles yield much higher ideal efficiencies than conventional effective temperatures predict, even when the light is unpolarized.

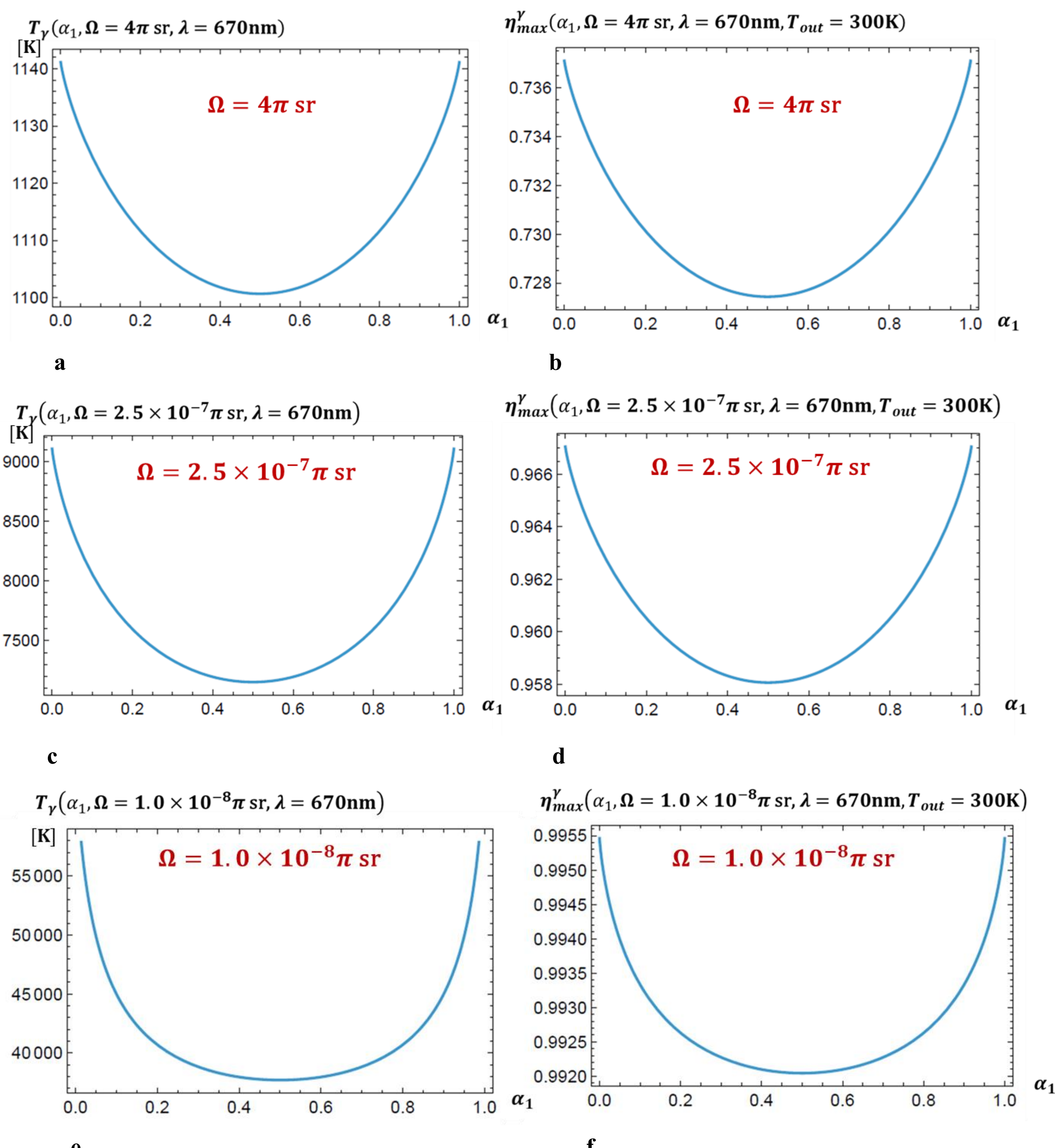

**Fig. 5. Dependence of the radiation temperature $T_\gamma(\alpha_1, \Omega, \lambda = 670nm)$ (left column) and the ideal efficiency $\eta^\gamma_{max}(\alpha_1, \Omega, \lambda = 670nm, T_{out} = 300K)$ (right column) on the polarization parameter $\alpha_1$ for the lights at three different solid angles $\Omega$ calculated by Eq.(4.18) : (a), (b) The case of $\Omega = 4\pi[sr]$** When $\alpha_1 = 1/2$, corresponding to unpolarized light, the results agree with the values predicted by conventional effective temperatures. **(c), (d) The case of $\Omega = 2.5 \times 10^{-7}\pi[sr]$, (e),(f) The**

**case of $\mathbf{\Omega = 1.0 \times 10^{-8}\pi[sr]}$** The latter two solid angles correspond to the conditions of regular lasers.

## 6. Ideal efficiency of photosynthesis on the earth considering the change in entropy due to the photochemical reaction

This section presents an analysis of the ideal efficiency of a terrestrial photosynthetic system under diluted solar irradiation, taking into account the change in entropy due to the photochemical reaction within the photosynthetic system and referring to a previous report by the present author [28].

The analysis is carried out in two steps in this study: (1) ignoring the change in entropy due to the photochemical reaction and (2) taking this change into account.

**(1)** $S_{in}^{\gamma D}(\lambda, \Omega = 4\pi)$ of the diluted solar radiation expressed in Eq. (5.6) can be rearranged as follows:

$$S_{in}^{\gamma D(1)}(\lambda, \Omega = 4\pi) = k_B \ln\left\{\frac{D^2}{R^2}\left(e^{hc/\lambda k_B T_{sun}} - 1\right) + 1\right\} N_{in}^{\gamma}(\lambda)$$

$$= \frac{hc}{T_{sun}\lambda} N_{in}^{\gamma}(\lambda) + k_B \left[\ln\left(\frac{D^2}{R^2}\right) + \ln\left\{1 - \left(1 - \frac{R^2}{D^2}\right) e^{-hc/\lambda k_B T_{sun}}\right\}\right] N_{in}^{\gamma}(\lambda). \quad (6.1)$$

The first term in Eq. (6.1) is the entropy of blackbody radiation at solar temperature $T_s$, and the second term is the increased entropy due to dilution as the radiation is transported from the Sun to the Earth, which becomes apparent by scattering in the molecules of the Earth's atmosphere. Applying Eq. (6.1) to Eq. (3.7) assuming $S_W = 0$ in Eq. (3.6), $\eta_{max}^{\gamma D}$ can be rearranged as

$$\eta_{max}^{\gamma D}(\lambda, \Omega = 4\pi, \widehat{n_D}(\lambda, T_S), S_W = 0)$$

$$= 1 - \frac{T_{out}}{hc/\lambda}\left[\frac{hc}{\lambda T_{sun}} + k_B \ln\left(\frac{D^2}{R^2}\right) + k_B \ln\left\{1 - \left(1 - \frac{R^2}{D^2}\right) e^{-hc/\lambda k_B T_{sun}}\right\}\right].$$

$$= 1 - \frac{T_{out}}{T_{sun}} - \eta^*(\lambda)$$

$$= \eta_C(T_{sun}) - \eta^*(\lambda), \quad (6.2)$$

where $\eta_C(T_{sun}) = 1 - \frac{T_{out}}{T_{sun}}$ is Carnot efficiency at solar temperature $T_{sun}$ and $\eta^*(\lambda) = \frac{T_{out}}{hc/\lambda} k_B \left[\ln\left(\frac{D^2}{R^2}\right) + \ln\left\{1 - \left(1 - \frac{R^2}{D^2}\right) e^{-hc/\lambda k_B T_{sun}}\right\}\right]$ is the reduced efficiency due to dilution. By substituting the required data, wavelength $\lambda = 670$ nm, solar temperature $T_{sun} = 5800$ K, distance from the Sun to the Earth $D = 1.50 \times 10^{11}$ m, radius of the Sun $R = 6.96 \times 10^{8}$ m, Boltzmann constant $k_B = 1.38 \times 10^{-23}$ J/K, and Planck constant $h = 6.63 \times 10^{-34}$ Js into Eq. (6.2), $\eta_C(T_{sun}) = 0.948$ and $\eta^* = 0.150$ were obtained and we consequently determined that $\eta_{max}^{\gamma D}(670\text{ nm}, \Omega = 4\pi, \widehat{n_D}(\lambda, T_{sun} = 5800\text{ K}), S_W = 0) = 0.798$. The corresponding effective temperature $T_D(\lambda = 670\text{ nm})$ was 1490 K.

(2) The analysis of the ideal efficiency of photosynthesis on the Earth's surface has been studied by the present author, taking into account the change in entropy due to the photochemical reaction. The photochemical reaction process of photosynthesis occurring in leaves exposed to monochromatic solar radiation of wavelength λ can be expressed by the following chemical equation,

$$6CO_2 + 6H_2O \rightarrow C_6H_{12}O_6 + 6O_2. \tag{6.3}$$

In the photosystem, $S_W$ in Eq. (3.6) is not zero for the ideal efficiency $\eta_{max}$. This $S_W$ is obtained from the change in entropy $\Delta_r S$ due to the photochemical reaction represented by Eq. (6.3). To calculate $\eta_{max}$ with $S_W \neq 0$ in Eq. (3.6) in a self-consistent way, the entropy $S_{in}^{\gamma}(\lambda, \Omega = 4\pi, \widehat{n_D}(\lambda))$ of the diluted solar radiation given by Eq. (6.1) is used.

The ideal efficiency with $S_W \neq 0$ can be obtained by the following procedure using Eq. (3.6).

$$\frac{W_{out}}{\boldsymbol{E}_{in}^{\gamma}(\lambda)} = \eta(\lambda, \widehat{n_D}(\lambda, T_{sun}), S_W) \leq 1 - \frac{T_{out}\left(\boldsymbol{S}_{in}^{\gamma(1)}(\lambda, \Omega=4\pi, \widehat{n_D}(\lambda)) - S_W\right)}{\boldsymbol{E}_{in}^{\gamma}(\lambda)}$$

$$= 1 - \frac{T_{out}S_{in}^{\gamma(1)}(\lambda, \Omega=4\pi, \widehat{n_D}(\lambda))}{E_{in}^{\gamma}(\lambda)} + \frac{T_{out}S_W}{E_{in}^{\gamma}(\lambda)}. \tag{6.4}$$

Further, by considering $1 - \frac{T_{out}S_{in}^{\gamma(1)}(\lambda, \Omega=4\pi, \widehat{n_D}(\lambda))}{E_{in}^{\gamma}(\lambda)} = \eta_C(T_{sun}) - \eta^*(\lambda)$ in Eq. (6.4), the following is obtained:

$$\frac{W_{out}}{\boldsymbol{E}_{in}^{\gamma}(\lambda)} \leq \eta_C(T_{sun}) - \eta^*(\lambda) + \frac{T_{out}S_W}{\boldsymbol{E}_{in}^{\gamma}(\lambda)} = \eta_C(T_{sun}) - \eta^*(\lambda) + \frac{T_{out}S_W}{W_{out}} \cdot \frac{W_{out}}{\boldsymbol{E}_{in}^{\gamma}(\lambda)}. \tag{6.5}$$

Solving this inequality for the energy efficiency $\eta(\lambda) = W_{out}/E_{in}(\lambda)$ gives the following inequality:

$$\eta^{\gamma}(\lambda) = \frac{W_{out}}{\boldsymbol{E}_{in}^{\gamma}(\lambda)} \leq \frac{\eta_C(T_{sun}) - \eta^*(\lambda)}{\left(1 - \frac{T_{out}S_W}{W_{out}}\right)}. \tag{6.6}$$

This inequality implies that

$$\tilde{\eta}_{max}^{\gamma}(\lambda, \Omega = 4\pi, \widehat{n_D}(\lambda, T_{sun}), S_W) = \frac{\eta_C(T_{sun}) - \eta^*(\lambda)}{\left(1 - \frac{T_{out}S_W}{W_{out}}\right)}, \tag{6.7}$$

where $W_{out}$ is the enthalpy of synthesized glucose and $S_W$ is obtained from the entropy change $\Delta_r S$

caused by the chemical reaction given by Eq. (6.3). Since these two quantities are only included as a ratio, their values per mole of glucose can be used. It should be noted that although the numerator on the right-hand side of Eq. (6.7) can be calculated using the effective temperature formulation, the denominator cannot be calculated by any method other than the energy-entropy flow analysis developed in this study. The value of this denominator as a correction factor can be calculated using the following data.

$$\begin{cases} S_W = \Delta_r S \ \text{(entropy change caused by the chemical reaction (6.3) with pressure correction)} \\ \quad = -663 \ \text{J/Kmol(glucose)} \\ W_{out} = H \ \text{(enthalpy of glucose)} = 2.8 \times 10^6 \ \text{J/mol(glucose)} \\ T_{out} = 300 \ \text{K (ambient temperature)} \end{cases} \tag{6.8}$$

The numerical value of $\left(1 - \frac{T_{out} S_W}{W_{out}}\right)$ obtained from these data was approximately $1.07$, and consequently the following formula was obtained:

$$\tilde{\eta}_{max}^{\gamma}(\lambda, \Omega = 4\pi, \widehat{n_D}(\lambda, T_{sun}), S_W)$$
$$= \frac{1}{1.07}(\eta_C(T_{sun}) - \eta^*)$$
$$= \frac{1}{1.07}\eta_{max}^{\gamma}(\lambda, \Omega = 4\pi, \widehat{n_D}(\lambda, T_{sun}), S_W = 0). \tag{6.9}$$

Where $\eta_{max}^{\gamma}(\lambda, \Omega = 4\pi, \widehat{n_D}(\lambda, T_{sun}), S_W = 0)$ is the maximum efficiency given by the conventional effective temperature $T_D^{\gamma}$. This implies that the theoretical maximum energy efficiency should be about 1/1.07 = 0.935 times smaller than that obtained when using only the effective temperature, as a result of correcting for the change in entropy due to the photochemical reaction. For example, according to Eqs. (6.2) and (6.9), at $\lambda$ = 670 nm and $T_S$ = 5800 K, the numerical value of $\tilde{\eta}_{max}^{\gamma}(\lambda, \Omega = 4\pi, \widehat{n_D}(\lambda, T_{sun}), S_W)$ was obtained as $\frac{1}{1.07} \times 0.798 = 0.745$.

The theoretical maximum energy efficiency $\eta_{max}$ of photosynthetically active radiation (PAR) with a wavelength in the range 400-700 nm absorbed by several types of terrestrial plants was calculated using the following formula given by Eqs. (6.10) and (6.11) below, based on an actual absorption spectrum. Consequently, the results of the calculations for these different types of plants indicated that the theoretical maximum energy efficiencies $\eta_{max}^{D}(\lambda{:}\, PAR, \Omega = 4\pi, \widehat{n_D}(\lambda, T_{sun}), S_W)$ of photosynthesis were within $\pm 2\%$ or 3% of 75%, depending on the type of plant [28]:

$$\tilde{\eta}^{\gamma}_{max}(\lambda: PAR, \Omega = 4\pi, \widehat{n_D}(\lambda, T_{sun}), S_W) = \frac{\eta_C(T_{sun}) - \eta^*(PAR)}{\left(1-\frac{T_{out}S_W}{W_{out}}\right)}, \quad (6.10)$$

where $\eta^*(PAR)$ is the reduction due to dilution caused by the expansion of sunlight and is given by the following formula:

$$\eta^*(PAR) = \frac{k_B T_{out}}{hc}\frac{1}{\sum_i \Delta N(\lambda_i)/\lambda_i}\sum_i \Delta N(\lambda_i)\left[\ln\left(\frac{D^2}{R^2}\right) + k_B \ln\left\{1-\left(1-\frac{R^2}{D^2}\right)e^{-hc/\lambda_i k_B T_{sun}}\right\}\right], \quad (6.11)$$

where the sum of wavelengths $\lambda_i$ is done in the PAR range (400–700 nm). The results of $\eta^*(PAR)$, calculated by fitting the actual absorption spectral data of some photosynthetic organisms to Eq. (6.11), and $\tilde{\eta}^{\gamma}_{max}(PAR)$ obtained by fitting these values [15] to Eq. (6.10), are shown in Table.

**Table. Comparison of the calculation results for various kinds of efficiencies in multiple**

| | *Phaseolus Vulgaris* | Several Plants Average | *Anabaena* |
|---|---|---|---|
| $\eta^{\gamma}_{actual}(PAR)$ | 0.230 | 0.164 | 0.182 |
| $\eta^{\gamma}_{C}$ | 0.948 | 0.948 | 0.948 |
| $\eta^*(PAR)$ | 0.134 | 0.133 | 0.131 |
| $\eta^{\gamma D}_{max}(PAR)$ | 0.814 | 0.815 | 0.817 |
| $\tilde{\eta}^{\gamma}_{max}(PAR)$ | 0.761 | 0.762 | 0.764 |

**photosynthetic organisms:** The actual efficiency $\eta^{\gamma}_{actual}(PAR)$, the Carnot efficiency $\eta^{\gamma}_{C}$, reduction parts $\eta^*(PAR)$ due to dilution effect, the maximum efficiency $\eta^{\gamma D}_{max}(PAR)$ without chemical reaction entropy (obtained even when using the effective temperature), and the final ideal efficiencies $\tilde{\eta}^{\gamma}_{max}(PAR)$ with chemical reaction entropy, for several photosynthetic organisms: These were calculated by fitting the actual absorption spectral data in the PAR range of several photosynthetic organisms [15] to Eqs. (6.10) and (6.11).

## 7. Photochemical transduction from radiation to a two-level pigment system in terms of entropy

This section considers a two-level pigment system consisting of the ground state pigment $P$ and the excited state pigment $P^*$,where light absorption induces a transition from $P$ to $P^*$. When such a pigment system is exposed to a certain intensity of light (a certain photon flux density per wavelength $\hat{n}_{\gamma}(\lambda)$ ), the entropy of radiation is transferred to the pigment system. In previous studies [9,12,30], the concentration ratio $[P^*]/[P]$was given as the following formula by substituting the effective temperature $T_{\gamma}(\lambda, \hat{n}(\lambda))$ for the Boltzmann-type factor,

$$\frac{[P^*]}{[P]} = \exp\left(-\frac{hc/\lambda}{k_B T_\gamma(\lambda,\hat{n}(\lambda))}\right). \quad (7.1)$$

Critical arguments have been made against this because non-equilibrium temperatures (i.e., not true temperatures) are automatically substituted into the Boltzmann-type factor equation without any basis [13,30]. In contrast, in this study, Eq. (7.1) was derived in terms of entropy analysis in the first-order evaluation based on quantum statistics, without recourse to radiation temperature such as $T_\gamma(\lambda,\hat{n}(\lambda))$. For this purpose, the change in entropy $\Delta S_p$ of this pigment system has to be formulated and matched with the imported entropy $S_{in}(\lambda,\Omega)$. In view of this, the following specific analysis is carried out.

According to the second law of thermodynamics, the increase in the entropy of the pigment system $\Delta S_p$ is not smaller than the absolute value of the decrease in radiation entropy $\Delta S_\gamma$. When both are equal, i.e., when the reversibility condition $\Delta S_\gamma + \Delta S_p = 0$ defined by entropy generation $S_g$=0 is satisfied, and assuming that $\Delta S_p$ is caused solely by an increase in the $n_{P^*}$ number ratio (number concentration) distribution of pigment molecules, the following discussion can be developed based on the formulation developed in this study.

Since a pigment system consists of Fermi particles, its entropy is obtained using quantum Fermi statistics as follows:

$$S_P = -k_B G\{f\ln f + (1-f)\ln(1-f)\}. \quad (7.2)$$

In molecular systems, because the particle number $N$ is sufficiently smaller than the number of quantum states $G$, i.e.,$f = \frac{N}{G} \ll 1$ is satisfied, $S_P$ expressed as Eq. (7.2) can be approximated as

$$S_P = -k_B G\{f\ln f - f\}. \quad (7.3)$$

Using Eq. (2.3) as $\Delta q^3 = V$, the distribution function $f = N/G$ can be expressed as

$$f = \frac{h^3 N}{\Delta p^3 V}. \quad (7.4)$$

By some calculations, after substituting Eq. (7.4) into Eq. (7.3), the following is obtained:

$$S_P(N) = -k_B N\ln\left(\frac{N}{V}\right) - k_B N\left\{\ln\left(\frac{h^3}{\Delta p^3}\right) - 1\right\}. \quad (7.5)$$

In the phenomenon considered here, only the concentration $N/V$ changes and the rest can be regarded as approximately constant, so the first term in equation (7.5) can be treated and analyzed as the substantial entropy of this phenomenon. Consequently, the entropy $S_c$=$-k_B N\ln(N/V)$=$-k_B N\ln C$, which is used in the analysis of chemical reactions and can be referred to as the concentration entropy ($C$ is a concentration value), can be derived.

Considering the additivity of entropy, the following formula is derived:

$$S_P(n_P, n_{P^*}) = S_P(n_P) + S_P(n_{P^*})$$

$$= -k_B n_P \ln\left(\frac{n_P}{V}\right) - k_B n_{P^*} \ln\left(\frac{n_{P^*}}{V}\right) - k_B N_P \left\{\ln\left(\frac{h^3}{\Delta p^3}\right) - 1\right\}, \quad (7.6)$$

where $n_{P^*}$ and $n_P$ are the number of excited state and ground state pigments, respectively, and $N_P$ is the total number of pigments. After simple calculations, $S_P(n_P, n_{P^*})$ is obtained as

$$S_P(n_P, n_{P^*}) = -k_B N_P \left\{\frac{n_{P^*}}{N_P} \ln\left(\frac{n_{P^*}}{N_P}\right) + \frac{n_P}{N_P} \ln\left(\frac{n_P}{N_P}\right)\right\} - k_B N_P \left\{\ln\left(\frac{N_P h^3}{V\Delta p^3}\right) - 1\right\}$$

$$= -k_B N_P \{p(n_P)\ln p(n_P) + p(n_{P^*}) p \ln(n_{P^*})\} - k_B N_P \left\{\ln\left(\frac{N_P h^3}{V\Delta p^3}\right) - 1\right\}, \quad (7.7)$$

where $p(n_P)$ and $p(n_{P^*})$ denote the ratios of $n_P$ and $n_{P^*}$ to $N_P$. In this discussion, it is assumed that the contribution from changes in the second term in Eq.(7.7) is negligible and only the contribution from changes in the first term is considered thereafter. The first term, which depends on $n_P$ and $n_{P^*}$ in Eq. (7.7), has the same form as that of the Shannon entropy, which is used as a measure of information. If $p(n_P) = n_P/N_P$ is expressed by $f_P$, $p(n_{P^*}) = n_{P^*}/N_P$ becomes $(1 - f_P)$ and the first term depending on $n_P$ and $n_{P^*}$ in Eq.(7.7) is transformed into

$$S_P(n_P) = -k_B N_P \{(1 - f_P)\ln(1 - f_P) + f_P \ln f_P\}. \quad (7.8)$$

This mathematical form is the same as that in Eq. (7.2).

Using equation (7.8), the first-order evaluation of the entropy change $\Delta S_p(n_P)$ due to the reduction $\Delta n_P (< 0)$ of the ground-state pigment molecules caused by photon absorption in the system is obtained as follows

$$\Delta S_P^{(1)}(n_P) = \frac{\partial S_p(n_P)}{\partial n_P} \Delta n_P$$

$$= \frac{\partial S_p(n_P)}{\partial f_P} \frac{1}{N_P} \Delta n_P$$

$$= k_B \ln\left(-1 + \frac{1}{f_P}\right) \Delta n_P \ . \quad (7.9)$$

By representing Eq. (7.9) in terms of $n_{P^*}$ and $n_P$, the following is obtained:

$$\Delta S_p^{(1)}(n_P) = k_B \ln\left(\frac{n_{P^*}}{n_P}\right) \Delta n_P = k_B \ln\left(\frac{[P^*]}{[P]}\right) \Delta n_P \quad (7.10)$$

Using Eq. (4.7), the formula for the entropy of a photon population (radiation) absorbed by the pigments can be obtained as follows:

$$S_{in}^{\gamma(1)}(\lambda,\Omega) = \frac{hc/\lambda}{T_\gamma(\lambda,\hat{n}(\lambda),\Omega)} N_{in}^{\gamma}(\lambda). \tag{7.11}$$

Thus, assuming that when radiation energy is absorbed by a pigment system, its entropy is also absorbed without generating additional entropy, i.e., assuming the reversibility condition, the absolute value of the decrease of the radiation entropy $\Delta S_\gamma$ is equal to the increase of the entropy of the system $\Delta S_p$, i.e., $\Delta S_\gamma + \Delta S_p = 0$. Then, $\Delta S_p(n_P) = S_{in}(\lambda,\Omega)$ is satisfied because $S_{in}(\lambda,\Omega)$ corresponds to $(-\Delta S_\gamma)$. Furthermore, the following equation is satisfied by considering Eqs. (7.10) and (7.11):

$$k_B \ln\left(\frac{[P^*]}{[P]}\right) \Delta n_P = \frac{hc/\lambda}{T_\gamma(\lambda,\hat{n}(\lambda),\Omega)} N_{in}^{\gamma}(\lambda). \tag{7.12}$$

The number of photons absorbed by a pigment system is equal to the number of pigments excited by photon absorption, because the process of absorbing photons into the pigment occurs by a one-to-one interaction between a photon and an electron in the pigment. Thus, $N_{in}(\lambda) = \Delta n_{P^*} = -\Delta n_P$ is satisfied, and the following equations are obtained from Eq. (7.12):

$$-k_B \ln\left(\frac{[P^*]}{[P]}\right) = \frac{\frac{hc}{\lambda}}{T_\gamma(\lambda,\hat{n}(\lambda),\Omega)}, \tag{7.13}$$

and from Eq. (4.7)

$$\frac{[P^*]}{[P]} = \exp\left(-\frac{hc/\lambda}{k_B T_\gamma(\lambda,\hat{n}(\lambda),\Omega)}\right) = \frac{\mathbf{1}}{\mathbf{1}+\frac{\mathbf{2\Omega c}}{\boldsymbol{\lambda^4 \hat{n}_\gamma(\lambda)}}}. \tag{7.14}$$

In the most general case including polarized light with polarization parameter $\alpha_1, \alpha_2$, the following general formula can be obtained from Eq. (4.18):

$$\frac{[P^*]}{[P]} = \exp\left(-\frac{hc/\lambda}{k_B T_\gamma(\lambda,\hat{n}(\lambda),\Omega,\alpha_1,\alpha_2)}\right) = \prod_{i=1}^{2}\left(\frac{1}{1+\frac{\Omega c}{\alpha_i \lambda^4 \hat{n}_\gamma(\lambda)}}\right)^{\alpha_i}. \tag{7.15}$$

Figure 6. shows the dependence of the concentration ratio $[P^*]/\ [P]$ on the polarization parameter $\alpha_1$ using Eq. (7.15) formulated in this study for light at three different solid angles of $\Omega = 4\pi[\mathrm{sr}]$, $\Omega = 2.5 \times 10^{-7}\pi[\mathrm{sr}]$, and $\Omega = 1.0 \times 10^{-8}\pi[\mathrm{sr}]$, respectively, at a wavelength of 670 nm. These results show that, like the ideal efficiency $\eta_{max}^{\gamma}$ shown in Figure 5, the dependence of the concentration ratio $[P^*]/\ [P]$ on the degree of polarization is smaller than its solid angle dependence. However, unlike the case of $\eta_{max}^{\gamma}$, $[P^*]/\ [P]$ for completely polarized light is found to be about twice that for unpolarized light, indicating a non-negligible difference.

In the case of unpolarized radiation (i.e., $\alpha_1 = \alpha_2 = 1/2$) with a solid angle of $4\pi$, $T_\gamma(\lambda,\hat{n}(\lambda),\alpha_1 =$ $\alpha_2 = 1/2, \Omega = 4\pi)$ in Eq. (7.15) coincides with Eq. (7.1) obtained by the conventional method based on the effective temperature. Eq. (7.1) has been questioned and criticized because this formula was derived automatically by applying the effective temperature $T_\gamma\left(\lambda, \hat{n}(\lambda)\right)$ to the Boltzmann-type factor even though it is not in equilibrium [9,12,30]. In contrast, according to the above considerations of this study under unpolarized and isotropic radiation, the condition to satisfy Eq. (7.1) is not an equilibrium radiation condition as in blackbody radiation, but a quasi-equilibrium condition between a radiation and a pigment molecular system, which condition guarantees the reversibility condition, i.e., the absence of entropy generation ($\Delta S_\gamma + \Delta S_p = 0$). The quasi-equilibrium condition for the radiation and pigment molecular system is expressed in terms of the applicability of first-order evaluation due to changes in photon and pigment numbers with respect to entropy change.

The above discussion and analysis is based on the assumption that when the photons in the radiation are absorbed by the pigment molecules in the ground state and become an excited state, all the entropy of the photon population is transferred to the $n^*$ number ratio (number concentration) distribution of the pigment molecules. Note that Eq. (7.12), Eq. (7.14) and Eq. (7.15) were derived only because of this basic premise. In the actual phenomenon of absorption of photons by the pigment molecules (internal electrons), there is not only energy transfer due to energy conservation, but also momentum transfer due to momentum conservation. Therefore, it is possible that the momentum contribution term, $-k_B N_P \ln(N_P h^3 / V \Delta p^3)$, in the expression for the entropy of the pigment molecule population in Eq. (7.7) is a non-negligible contribution to the actual entropy transfer from the photon population to the pigment molecule population due to the absorption of photons by the pigment molecules (internal electrons). This is an issue to be addressed in the future.

It should be noted that the above discussion and analysis is based on the applicability of the first-order evaluation by $\Delta N/N$, which is guaranteed by the quasi-equilibrium conditions between the radiation and the system. This is discussed and analyzed in detail in Part 2 of this study [6].

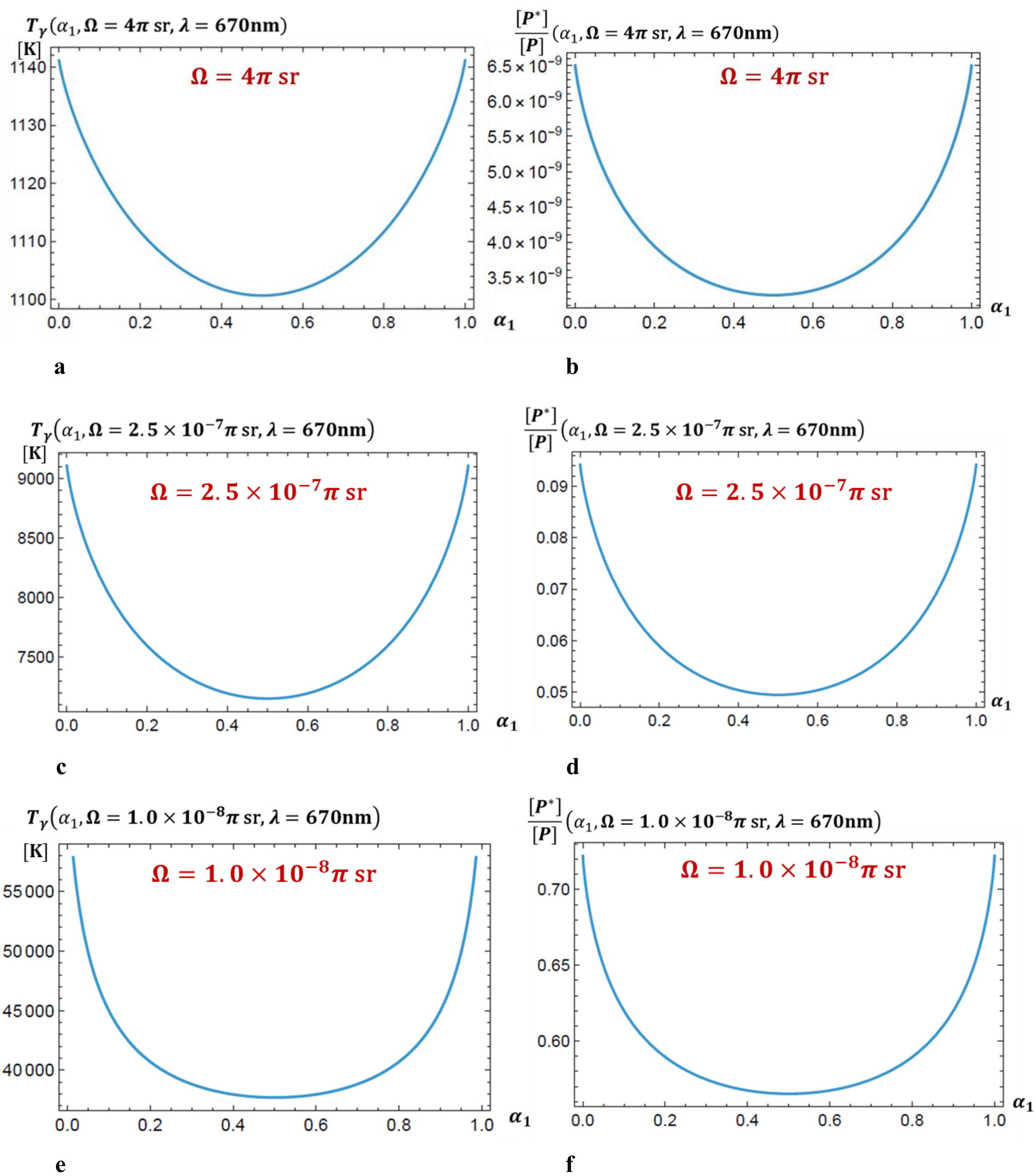

**Fig. 6. Dependence of the radiation temperature $T_\gamma(\alpha_1, \Omega, \lambda = 670\ \mathrm{nm})$ (left column) and the concentration ratio $[P^*]/\ [P]$ (right column) on the polarization parameter $\alpha_1$ for the lights at three different solid angles calculated by Eq.(7.15) : (a),(b) The case of $\Omega = 4\pi[\mathrm{sr}]$** When $\alpha_1 = 1/2$, corresponding to unpolarized light, the results agree with the values predicted by conventional effective temperatures. **(c),(d) The case of $\Omega = 2.5 \times 10^{-7}\pi[\mathrm{sr}]$, (e),(f) The case of $\Omega = 1.0 \times 10^{-8}\pi[\mathrm{sr}]$** These two solid angles correspond to the conditions of regular lasers.

## 8. Equilibrium and Reversibility

The main criticism and question raised by previous studies against the use of the effective temperature $T_\gamma$ in the analysis of light-powered system, such as photosynthesis, using the sunlight at the surface of Earth that is diluted on its way from the Sun to the Earth, was that the evaluations of $\eta_{max}\left(T_\gamma(\lambda)\right)$, and the Boltzmann-type factor $\exp\left(-\Delta E/k_B T_\gamma(\lambda)\right)$ used as a formula giving the concentration ratio of excited-to-ground state pigment-molecules by such a radiation temperature are incorrect because such diluted light is not in thermal equilibrium with a blackbody (i.e., it is not blackbody radiation). In reality, as this study has shown, diluted sunlight reaching the surface of Earth does not satisfy the thermal equilibrium condition as a blackbody, but if it satisfies the quasi-equilibrium condition between the radiation and the system (a first-order evaluable condition by the number of photons absorbed) as it enters a system, then the analysis using the radiation temperature $T_\gamma(\lambda)$ was consequently not incorrect. The essential reason for the agreement between Eq. (4.8) obtained by the theoretical method in this study and Eq. (4.10) obtained by the conventional method based on the extension of the blackbody radiation temperature, is discussed here.

### 8.1. Reversible condition

Carnot engine is an ideal heat-powered system consisting of reversible processes without entropy generation along the way; this reversible condition gives the theoretical maximum energy efficiency (e.g. of heat engines and light-powered systems), as formulated by the analysis of entropy and energy flows in Section 3. However, for Carnot efficiency of heat-powered systems, the pseudo-Carnot efficiency of light-powered systems, etc., not only reversible conditions but also equilibrium conditions, strictly speaking quasi-equilibrium conditions are required. When the energy source is radiation, the equilibrium condition is not the equilibrium condition with the blackbody (blackbody radiation condition), where the temperature can be defined, but the equilibrium condition between the radiation and the system. This would be something that has been overlooked in previous studies. The first-order evaluable condition $\Delta T/T \ll 1$, which gives the ideal efficiency of the heat-powered system, i.e., the Carnot efficiency, provides not only the quasi-equilibrium condition but also one of the conditions that guarantees reversibility.

### 8.2. Equilibrium condition

Equilibrium is defined as a state in which all macroscopic changes have ceased. According to the ~~above~~ definition of equilibrium state, the state in which the microscopic traffic of particles and energy between the two systems is balanced and the macroscopic traffic has ceased, is an equilibrium state. Therefore, the Carnot engine has been assumed to be in equilibrium in two respects. One is the thermal equilibrium state of the energy source (heat bath) with a definite temperature $T$, and the other is the equilibrium state as a state in which thermal energy has stopped flowing back and forth between the energy source and the system. However, if the equilibrium between the heat bath and the system were perfect, the flow of thermal energy from the heat bath, which is the energy source, to the system would cease and no power could be extracted. Therefore, in the Carnot engine, the process of thermal energy flow between the heat bath and the system has been assumed to be not in perfect equilibrium but almost in equilibrium between them. This is exactly the isothermal process in a Carnot engine, and this assumption is called quasi-static process. This assumption implies the condition that the thermal change $\Delta T$ associated with the heat transfer is infinitesimal, i.e., the first-order evaluable condition $\Delta T/T \ll 1$. In fact, this first-order evaluable condition is also implicitly assumed in many formulae underlying classical thermodynamics, including the Clausius formula given by $dS = d'Q/T$. This formula is derived by the first-order evaluation by $\Delta T/T$ as Eq. (3.4) under the constant volume, based on the following formula for the entropy (per mole) of an ideal gas obtained from the Sackur-Tetrode equation in thermo-statistical mechanics.

$$S = C_V \ln T + R \ln V, \tag{8.1}$$

where $C_V$ is the heat capacity per mole and $R$ is the gas constant.

**8.3. The essential reason for the agreement between two methods**

In the conventional method, $\hat{n}_{BB}(\lambda, T)$ in Eq. (4.9) for blackbody radiation is replaced by an arbitrary $\hat{n}(\lambda)$ and solved for $T$ to obtain the effective temperature $T_D$ as $T = T_D(\lambda)$. This method is used to obtain the effective temperature, which is employed in radiation thermodynamics for applications such as lasing and fluorescent cooling [31]. To understand this method, it helps to review the equation for blackbody radiation in the following way. Blackbody radiation in thermal equilibrium has a maximum entropy under the condition that the total energy of the photon ensemble and the surrounding wall, as a heat bath, is constant, and its time variation has ceased, so that $\partial S^\gamma/\partial t = 0$ holds. Based on this point, the analysis can be done as follows.

$$\begin{aligned}\frac{\partial S^\gamma}{\partial t} &= \sum_i \frac{\partial S^\gamma}{\partial N^\gamma(\lambda_i)} \frac{dN^\gamma(\lambda_i)}{dt} \\ &= \sum_i \frac{\partial S^\gamma}{\partial f(\lambda_i)} \frac{\partial f(\lambda_i)}{\partial N^\gamma(\lambda_i)} \frac{dN^\gamma(\lambda_i)}{dt} \\ &= \sum_i k_B \ln\left(1 + \frac{1}{f(\lambda_i)}\right) \frac{dN^\gamma(\lambda_i)}{dt}\end{aligned} \tag{8.2}$$

Since the total energy of the photon ensemble is constant,

$$\sum_i \frac{hc}{\lambda_i}\frac{dN^{\gamma}(\lambda_i)}{dt} = 0 \tag{8.3}$$

holds. From Eqs. (8.2) and (8.3), the dimensional consideration and the thermodynamic findings, the following formula is obtained,

$$\frac{\partial S^{\gamma}}{\partial N(\lambda)} = k_B \ln\left(1+\frac{1}{f(\lambda)}\right) = \frac{1}{T}\frac{hc}{\lambda}. \tag{8.4}$$

Solving Eq. (8.4) for $f(\lambda)$, we obtain the distribution function of blackbody radiation, $f(\lambda) = 1/\left(e^{hc/\lambda k_B T}-1\right)$, and solving Eq. (8.4) for $T$ under the given photon flux density $n(\lambda)$, we obtain the radiation temperature as follows.

$$T_{\gamma}(\lambda) = \frac{hc/\lambda}{\frac{\partial S^{\gamma}}{\partial N(\lambda)}} = \frac{hc/\lambda}{k_B \ln\left(1+\frac{1}{f(\lambda)}\right)}. \tag{8.5}$$

Substituting $f(\lambda, \Omega = 4\pi)$ in Eq. (2.3) into Eq. (8.5) using the photon flux density per wavelength $\hat{n}(\lambda) = (1/4)\rho_N(\lambda, T)c/\Delta\lambda$ and adding some expansion, Eq. (4.10) is obtained.

The $\partial S/n(\lambda)$ in Eq. (8.5) results from the equilibrium conditions of blackbody radiation. The $\partial S/N^{\gamma}(\lambda)$ in Eq. (4.2), on the other hand, results from the first-order evaluable condition in the following process of deriving the maximum efficiency $\eta^{\gamma}_{max}(\lambda)$.
In the process of expanding the formula of

$$\eta^{\gamma}_{max}(\lambda) = 1 - T_{out}S_{in}(\lambda)/E_{in}(\lambda) = 1 - T_{out}\frac{\frac{\partial S^{\gamma}}{\partial n(\lambda)}(-\Delta N(\lambda))}{\frac{hc}{\lambda}(-\Delta N(\lambda))} = 1 - T_{out}\frac{\frac{\partial S^{\gamma}}{\partial N(\lambda)}}{\frac{hc}{\lambda}} = 1 - \frac{T_{out}}{T_{\gamma}(\lambda)}, \tag{8.6}$$

$\partial S^{\gamma}/\partial N(\lambda)\cdot\left(-\Delta N(\lambda)\right)$ has been used as the first-order evaluation of $S^{\gamma}_{in}(\lambda)$ by $\Delta N(\lambda)$. Note that this first-order evaluation is only valid under the quasi-equilibrium condition between a photon ensemble(radiation) and a system. This equilibrium condition is satisfied only when the photon number absorption rate $|\varepsilon| = -\Delta N(\lambda)/N(\lambda)$ of the system is sufficiently small ($|\varepsilon| \ll 1$). Otherwise the exact formulation of $\eta^{\gamma}_{max}(\lambda, |\varepsilon|)$ should be correctly constructed. Its formulation is shown and analyzed in detail in Part 2 of this study [6].

As described above, the essential reason for the agreement between Eq. (4.10) obtained by the conventional method and Eq. (4.7) obtained by the method in this study can be understood as that both were obtained using $\partial S^{\gamma}/\partial N^{\gamma}(\lambda)$, based on equilibrium conditions between a photon ensemble (radiation) and a blackbody in the former, and between a photon ensemble (radiation) and a light-powered system in the latter. However, it is very important to note that the more general and realistic theoretical maximum energy efficiency of the light-powered system with finite light absorption rate cannot be formulated using the equilibrium condition, but only using the reversible condition within the system, as be formulated in Part 2 of this study [6].

## 9. Conclusions and Future Applications

The ideal efficiency (theoretical maximum efficiency $\eta_{max}$) for a photosynthetic system with arbitrary photon flux density $n_\gamma(\lambda)$, solid angle $\Omega$, and polarization $P$ has been formulated using energy and entropy flow analysis. The radiation temperature of monochromatic light, derived using the first-order evaluable condition of photon number change, which is a quasi-equilibrium condition between the radiation and the system, was found to be consistent with the conventional radiation temperature for a given light intensity when the $\Omega$ and $P$ of the radiation are 4π and 0, respectively. The method is then used in this paper to perform a number of analyses that were not possible using conventional methods. This formulation has applicability to practical analyses such as the following (1) to (4) that were not possible with conventional radiation temperature, including those induced by Landau-Lifshitz [24], which was obtained as an extension of the temperature of blackbody radiation.

(1) An analysis of the ideal efficiency under first-order evaluable conditions for light of arbitrary polarization has been presented and formulated in Eq. (4.17), from which the radiation temperature including polarization is formulated in Eq. (4.18). In these equations, the degree of polarization is denoted by the polarization parameter αi $\alpha_i$ $(i = 1,2)$.

(2) An analysis of the ideal efficiency of the terrestrial photosynthetic system under diluted solar irradiation, taking into account the change in entropy due to the photochemical reaction, is presented.

(3) Analyses of physical quantities such as the ideal efficiency and the ratio of excited-state pigment molecules to ground-state molecules under irradiation with arbitrary solid angle and degree of polarization were presented. They were formulated as Eq. (4.17) and Eq. (7.15), respectively.

(4) Analysis of physical quantities such as $\eta_{max}^{\gamma}$ of photosynthesis and photovoltaics by sunlight scattered through the atmosphere under various meteorological and geographical conditions is also possible with the formulation of this study, and research into this is currently underway.

For analysis (4), the formulation developed in this study can be used to formulate a general analysis that also takes into account the wavelength, zenith angle, azimuthal angle, and polarization dependence of the scattered light. An overview of the analysis is given below:

$$\eta_{max}^{\gamma} = 1 - \frac{T_{out} S_{in}^{\gamma}}{E_{in}^{\gamma}}, \tag{9.1}$$

where $E_{in}^{\gamma}$ and $S_{in}^{\gamma}$ are given by the following formulae, respectively:

$$E_{in}^{\gamma} = \sum_{i.j,k} \frac{hc}{\lambda_i} |\varepsilon(\lambda_i)|\, \hat{\hat{n}}_{\gamma}(\lambda_i, \theta_j, \varphi_k) \Delta\lambda 2\pi \sin\theta_j \Delta\theta\Delta\varphi \tag{9.2}$$

$$S_{in}^{\gamma} = k_B \sum_{\ell=1}^{2} \sum_{i.j,k} \alpha_{\ell}(\lambda_i, \theta_j, \varphi_k) \ln\left\{1 + \frac{4\pi c}{2\alpha_{\ell}(\lambda_i,\theta_j,\varphi_k)\hat{\boldsymbol{n}}_{\gamma}(\boldsymbol{\lambda}_{i,},\theta_j,\varphi_k)}\right\}$$

$$\cdot \{1 + \gamma(f(\lambda_i), \varepsilon(\lambda_i))\} |\varepsilon(\lambda_i)|\, \hat{n}_{\gamma}(\lambda_i, \theta_j, \varphi_k) \Delta\lambda 2\pi \sin\theta_j \Delta\theta\Delta\varphi, \qquad (9.3)$$

where $\lambda_i$, $\theta_j$ and $\varphi_k$ are the wavelength, zenith angle, and azimuth angle, respectively, discretized to take the sum instead of the integral, and $\hat{n}_{\gamma}(\lambda_i, \theta_j, \varphi_k)$ is the photon flux density per unit solid angle and unit wavelength interval with wavelengths $\lambda_i$ absorbed by a light-powered system , which is represented as a leaf in Fig.7. A theoretical analysis based on this formulation considering the entropy generation of solar radiation due to in the Earth's atmosphere, is being carried out by the author and collaborators, with reference to previous studies [23,32–35]. Such a precise analysis of the entropy of terrestrial solar radiation is expected to contribute to the local estimation of the ideal efficiency of agricultural products and solar power generation. The general formulations developed in this paper should contribute to such research as well.

As stated in a paper [36] celebrating Light Year 2015, future research on light will require not only an analysis of the energy of light, but also a detailed analysis of its entropy. The study developed in this paper will contribute to the future research in the following interdisciplinary applications [36]:

(1) To define and determine the second law photosynthetically active radiation (PAR) region efficiency.

(2) To measure and theoretically study entropy production in the Earth's atmosphere.

(3) To show how the evolution of human vision was driven by the entropy content of radiation.

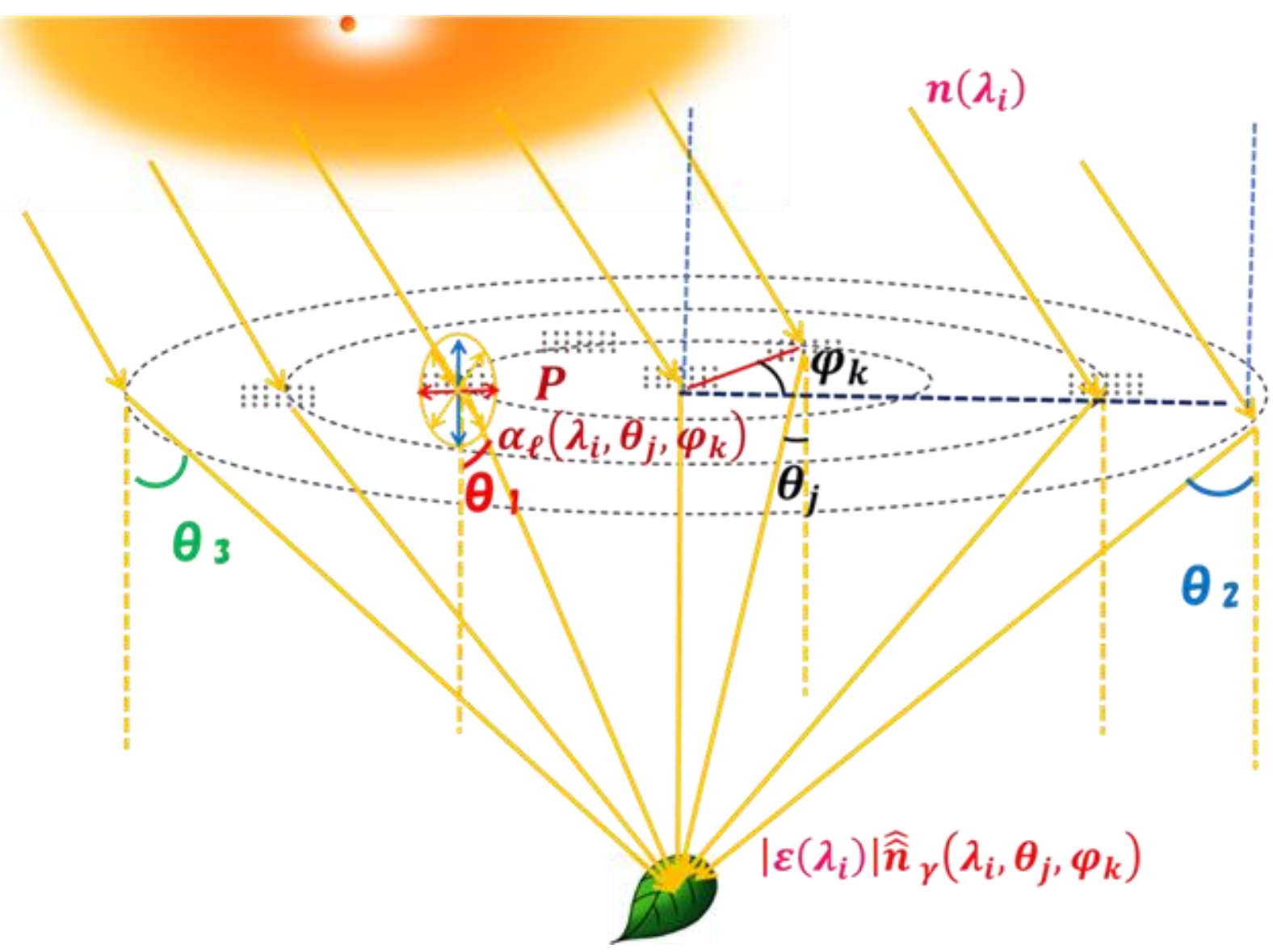

**Fig. 7. Sunlight incident on a light-powered system (a plant leaf) after being scattered by air molecules, aerosols, and cloud particles in the atmosphere:** This picture is depicted with a zenith angle $\theta_j$ , azimuth angle $\varphi_k$, and polarization parameter $\alpha_{\ell}(\lambda_i, \theta_j, \varphi_k)(\ell = 1,2)$.

**Glossary**

LHC: Light Harvesting Complex

PAR: Photosynthetic Active Radiation

*Light-powered* system: In this study, this is defined not only as a system that outputs electrical energy, such as solar power generation, photovoltaics and solar cells, etc., but also as a system from which any kind of physical work is extracted using light energy, including natural and artificial photosynthesis

**Data availability**

No database was used and no new data were generated for this article.

**Funding**

This research received no external funding.

**Conflicts of Interest**

The author declares no conflicts of interest.

## ACKNOWLEDGMENTS

I would like to acknowledge Prof. Ayumi Tanaka and Prof. Iwao. Yamazki for the useful discussions on the photosynthesis systems and optical science, respectively, Prof. Yousuke Ohyama for the useful mathematical discussions, Dr. Nobuki Maeda and Prof. Takashi Uchida and Prof. Shigenori Tanaka for the useful physical discussions.

**Appendix A**

Here is the proof that $S(\lambda,\Omega,\alpha_1,\alpha_2)$, given by Eqs. (2.4) and (2.5), increases monotonically with the degree of polarization $P$ is presented.

$$S(\lambda,\Omega,\alpha_1,\alpha_2)$$

$$= k_B \frac{G(\lambda,\Omega)}{2} \sum_{i=1}^{2} \left\{ \left(1+2\alpha_i f(\lambda,\Omega)\right) \ln\left(1+2\alpha_i f(\lambda,\Omega)\right) - 2\alpha_i f(\lambda,\Omega) \ln\left(2\alpha_i f(\lambda,\Omega)\right) \right\}.$$

As $P$ is equal to $|\alpha_1-\alpha_2|$, the monotonic increase can be proven by showing that the following conditions are satisfied:

(1) $\frac{d}{d\alpha_1} S(\lambda,\Omega,\alpha_1,\alpha_2) > 0 \quad (0 \leq \alpha_1 < \frac{1}{2} < \alpha_2)$,

(2) $\frac{d}{d\alpha_1} S(\lambda, \Omega, \alpha_1, \alpha_2) = 0 \quad (\alpha_1 = \alpha_2 = \frac{1}{2})$,

(3) $\frac{d}{d\alpha_1} S(\lambda, \Omega, \alpha_1, \alpha_2) < 0 \quad (0 \leq \alpha_2 < \frac{1}{2} < \alpha_1)$.

Following certain calculations, the subsequent result is obtained:

$$\frac{d}{d\alpha_1} S(\lambda, \Omega, \alpha_1, \alpha_2)$$

$$= \left(\frac{\partial}{\partial\alpha_1} - \frac{\partial}{\partial\alpha_2}\right) S(\lambda, \Omega, \alpha_1, \alpha_2)$$

$$= k_B N(\lambda) \ln\left\{\frac{\alpha_2 + 2\alpha_1\alpha_2 f(\lambda, \Omega)}{\alpha_1 + 2\alpha_1\alpha_2 f(\lambda, \Omega)}\right\}. \quad \text{(A.1)}$$

Eq. (A.1) obviously satisfies the conditions (1), (2), and (3). Eq. (4.14), which is the entropy imported into a light-powered system by a monochromatic light, has been formulated including the polarization freedom $\alpha_1, \alpha_2$ as

$$S_{in}^{\gamma(1)}(\lambda, \Omega, \alpha_1, \alpha_2) = k_B \sum_{i=1}^{2} \alpha_i \ln\left(1 + \frac{1}{2\alpha_i f(\lambda, \Omega)}\right) \Delta N(\lambda). \quad \text{(A.2)}$$

After some calculations, we can also see that Eq. (A.2) increases monotonically with increasing degree of polarization $P$ as follows:

$$\frac{d}{d\alpha_1} S_{in}(\lambda, \Omega, \alpha_1, \alpha_2) = \left(\frac{\partial}{\partial\alpha_1} - \frac{\partial}{\partial\alpha_2}\right) k_B \sum_{i=1}^{2} \alpha_i \ln\left(1 + \frac{1}{2\alpha_i f(\lambda, \Omega)}\right) \Delta N(\lambda)$$

$$= k_B \left[\frac{\partial}{\partial\alpha_1}\left\{\alpha_1 \ln\left(1 + \frac{1}{2\alpha_1 f(\lambda, \Omega)}\right)\right\} - (\alpha_1 \leftrightarrow \alpha_2)\right] \Delta N(\lambda)$$

$$= k_B \left[\left\{\ln\left(1 + \frac{1}{2\alpha_1 f(\lambda, \Omega)}\right) - \frac{1}{1 + 2\alpha_1 f(\lambda, \Omega)}\right\} - (\alpha_1 \leftrightarrow \alpha_2)\right] \Delta N(\lambda), \quad \text{(A.3)}$$

where the second term $(\alpha_1 \leftrightarrow \alpha_2)$ represents one exchanged between $\alpha_1$ and $\alpha_2$ in the second term. From Eq. (A.3) we can see that $\frac{d}{d\alpha_1} S_{in}(\lambda, \Omega) = 0$ at $\alpha_1 = \alpha_2 = \frac{1}{2}$, and after some calculation, we obtain

$$\frac{d^2}{d{\alpha_1}^2} S_{in}(\lambda, \Omega, \alpha_1, \alpha_2) = -k_B \left\{\frac{1}{\left(1 + 2\alpha_1 f(\lambda, \Omega)\right)^2} + \frac{1}{\left(1 + 2\alpha_2 f(\lambda, \Omega)\right)^2}\right\} \Delta N(\lambda) < 0. \quad \text{(A.4)}$$

These results indicate that $S_{in}(\lambda, \Omega)$ is maximized at $\alpha_1 = \alpha_2 = \frac{1}{2}$ (unpolarized), and minimized at $\alpha_1 = 1, \alpha_2 = 0$ or $\alpha_1 = 0, \alpha_2 = 1$ (completely polarized). Thus, the theoretical maximum energy efficiency $\eta_{max}(\lambda, \Omega, \widehat{n_\gamma}(\lambda), \alpha_1, \alpha_2)$ given by Eq. (4.17) and the radiation temperature given by Eq. (4.18) decrease monotonically with the degree of polarization $P$.

Moreover, its maximum value increases monotonically with decreasing distribution function $f$.

**Appendix B. The relationship between the radiation entropy $S(\lambda, \Omega)$ and the radiation energy $E(\lambda, \Omega)$**

In Section 4, we formulated the calculation of the radiation entropy $S_{in}(\lambda, \Omega)$ imported into a powered system dependent on the imported radiation energy $E_{in}(\lambda, \Omega)$.

The quantity required to calculate the theoretical maximum energy efficiency of a light-powered system, including natural photosynthesis, is not the absolute value of entropy contained in radiation as the energy source of a light-powered system. Rather, it is a decreasing value of radiation entropy accompanying its energy transfer to a system. When the radiation serving as the energy source is in thermal equilibrium, i.e., blackbody radiation,

$$\frac{du(T)}{T} = ds(T) \tag{B.1}$$

holds not only for white light including the entire spectrum but also for monochromatic radiation.

This property was derived as follows. The infinitesimal change $ds(\lambda, T)$ in the entropy of monochromatic radiation with the wavelength $\lambda$ responding to change in temperature $dT$ can be expressed as

$$\begin{aligned} ds(\lambda, T) &= \frac{\partial}{\partial T} s(\lambda, T) dT \\ &= \frac{\partial s(\lambda, T)}{\partial f} \frac{\partial f}{\partial T} dT \\ &= k_B g(\lambda) \ln\left(1 + \frac{1}{f(\lambda, T)}\right) \frac{\partial f(\lambda, T)}{\partial T} dT. \end{aligned} \tag{B.2}$$

From Eq. (B.2), the formula of Eq. (B.1) for the monochromatic radiation with the wavelength $\lambda$ can be obtained as

$$\begin{aligned} \frac{ds(\lambda, T)}{du(\lambda, T)} &= \frac{k_B g(\lambda) \ln\left(1 + \frac{1}{f(\lambda, T)}\right) \frac{\partial f(\lambda, T)}{\partial T} dT}{(hc/\lambda) g(\lambda) \frac{\partial}{\partial T} f(\lambda, T) dT} \\ &= \frac{k_B \ln\left(1 + \frac{1}{f(\lambda, T)}\right)}{hc/\lambda} \end{aligned}$$

$$= \frac{k_B \ln\left(1+e^{hc/\lambda k_B T}-1\right)}{hc/\lambda}$$

$$= \frac{1}{T}, \qquad \text{(B.3)}$$

where $f(\lambda, T) = \left(e^{hc/\lambda k_B T} - 1\right)^{-1}$ is used.

Even when the radiation as energy source is not in thermal equilibrium, using Eq. (4.7), the same formula as Eq. (B.3) is obtained in terms of $T_\gamma(\lambda, \hat{n}(\lambda), \Omega)$ having temperature dimensions:

$$\frac{1}{T_\gamma(\lambda,\hat{n}(\lambda),\Omega)} = \frac{S_{in}^{\gamma(1)}(\lambda,\Omega)}{E_{in}^{\gamma}(\lambda)} = \frac{(ds(\lambda,\Omega)/dN(\lambda))dN(\lambda)}{(du(\lambda)/dN(\lambda))dN(\lambda)} = \frac{ds(\lambda,\Omega)}{du(\lambda)}. \qquad \text{(B.4)}$$

When the solid angle is 4π, $T_\gamma(\lambda, \hat{n}(\lambda), \Omega)$ in Eq. (B.4) becomes $T_\gamma(\lambda, \hat{n}(\lambda), \Omega = 4\pi)$, which is the radiation temperature, often called the effective temperature $T_D(\lambda)$.

**Appendix C**

The entropy of terrestrial solar radiation, diluted from the Sun to the Earth before entering the Earth's atmosphere and being scattered in it, does not increase and remains unchanged from the original value, because the increase in the entropy in configuration space (configuration-space entropy) is completely cancelled by the decrease in the entropy in momentum space (momentum-space entropy) due to the decrease in the solid angle $\Omega$. A specific and rigorous proof of this fact by the author, which is guaranteed by Liouville's theorem only generally, is given in this Appendix using elementary geometry.

The proof is given by showing that the distribution function $f(\lambda, t) = N(\lambda, t)/G\big(\lambda, \Omega(t), V(t)\big)$ of a photon ensemble is constant over time. In this appendix the proof is given by differential analysis.

In Fig. 8a, three distances $R, a,$ and $r$ and three angles $\theta, \phi,$ and $\phi + \theta$ are defined as follows. The distance $R$ is the radius of the Sun, $a$ (the line length AB in Fig. 8) is the traveling distance of a photon emitted towards B located at an angle $\theta$ from the center (O) of the Sun within time interval $\Delta t$, i.e., $a = c\Delta t$, where $c$ is the velocity of light, and $r$ (the line length OB) is the distance from the center of the Sun to the point of arrival B of the emitted photon. The angle $\phi + \theta$ is the zenith angle of the emitted photon on the Sun's surface, and $\phi$ is its zenith angle on the surface point B. Fig. 8b shows the solar sphere and the photosphere surface, whose temperature is about 5800 K, with a shell thickness $\Delta\ell$ as a focus and the corresponding spherical shell thickness $\Delta\ell'(\theta)$ of an emitted photon ensemble on the surface point B. Based on these considerations, the proof is carried out simultaneously by differential analysis for a photon ensemble with a given wavelength λ. From the conservation of photon numbers it suffices to consider that the distribution function in the differential phase space $f(\lambda, d\Omega, dV) = N(\lambda, d\Omega, dV)/G(\lambda, d\Omega, dV)$, which becomes the following general formula using Eq. (2.3),

$$f(\lambda,t) = N(\lambda)/\left(\frac{2d\Omega(t)dV(t)}{\lambda^4}\Delta\lambda\right) \tag{C.1}$$

is constant. The consideration and analysis is done in the following steps

(1) Differential volume of configuration-space $dV$

The width of a concentric spherical shell $d\ell$, $d\ell'$ is given by the differential length $dR$, $dr$, respectively (**Fig. 8**b). From spherical symmetry, we can assume that $dV_S = 4\pi R^2 dR$, $dV_E = 4\pi r^2 dr$.

(2) Differential volume of momentum-space $2p^2 dp d\Omega = \frac{2\Delta\lambda d\Omega(t)}{\lambda^4}$

Since the solid angle $\Omega$ is generally given by the formula $\Omega = 2\pi(1 - cos\theta)$ using the half-apex angle $\theta$, it can be seen that the differential solid angles $d\Omega_S$ on the Sun's surface and $d\Omega_E$ on the Earth are represented by $2\pi sin(\phi+\theta)d(\phi+\theta)$ and $2\pi sin\phi d\phi$, respectively.

An elementary geometric analysis yields the following equations (Fig. 8 a, b):

$$Rsin(\phi+\theta) = r\sin\phi \tag{C.2}$$

$$R\sin\theta = a\sin\phi \tag{C.3}$$

$$a\cos\phi + R\cos\theta = r \tag{C.4}$$

To make a comparison of an ensemble of photons that have passed the same time from emission, fixing $a = ct$ (photon travel distance) and performing total differentiation by $R, r, \phi, \theta$ and rearranging the formula with $dR, dr, d(\phi+\theta), d\phi$, the following differential formulae are obtained. From (C.2), (C.3):

$$\sin(\phi+\theta)dR + R\cos(\phi+\theta)d(\phi+\theta) = \sin\phi dr + r\cos\phi d\phi, \tag{C.5}$$

$$\sin\theta dR + R\cos\theta d(\phi+\theta) = (a\cos\phi + R\cos\theta)d\phi \tag{C.6}$$

And, (C.6) becomes the following formula using (C.4),

$$\sin\theta dR + R\cos\theta d(\phi+\theta) = r d\phi. \tag{C.7}$$

After (C.7)×$\cos(\phi+\theta)$ − (C.5)×$\cos\theta$ and some slight calculation with Sine addition theorem,

$$dR = \cos\theta dr + r\sin\theta d\phi \tag{C.8}$$

is obtained. By substituting (C.8) into (C.7), the following formula is obtained,

$$d(\phi+\theta) = -\sin\theta/R dr + r/R\cos\theta d\phi. \tag{C.9}$$

The formulae (C.8) and (C.9) are represented by the following compact matrix formula,

$$\begin{pmatrix} dR \\ d(\phi+\theta) \end{pmatrix} = \begin{pmatrix} \cos\theta & r\sin\theta \\ -\sin\theta/R & r\cos\theta/R \end{pmatrix} \begin{pmatrix} dr \\ d\phi \end{pmatrix}. \tag{C.10}$$

Since the determinant of the matrix on the right-hand side of (C.10) is a Jacobian, we can use its absolute value to obtain the following equation, which gives the relationship between the two area elements.

$$\begin{aligned} dRd(\phi+\theta) &= \det\begin{pmatrix} \cos\theta & r\sin\theta \\ -\sin\theta/R & r\cos\theta/R \end{pmatrix} drd\phi \\ &= r/R drd\phi \end{aligned}$$

As a result,

$$\frac{dRd(\phi+\theta)}{drd\phi} = r/R \tag{C.11}$$

is obtained. From the formulae $G_{Earth}(\lambda) = \frac{2d\Omega_E dV_E}{\lambda^4}\Delta\lambda$ and $G_{Sun}(\lambda) = \frac{2d\Omega_S dV_S}{\lambda^4}\Delta\lambda$, the following formula is obtained,

$$\begin{aligned} \frac{G_{Sun}(\lambda)}{G_{Earth}(\lambda)} &= \frac{d\Omega_S dV_S}{d\Omega_E dV_E} \\ &= \frac{2\pi sin(\phi+\theta)d(\phi+\theta)\, 4\pi R^2 dR}{2\pi sin\phi d\phi 4\pi r^2 dr} \\ &= \frac{sin(\phi+\theta)}{sin\phi}\cdot\frac{R^2}{r^2}\cdot\frac{dRd(\phi+\theta)}{drd\phi} \\ &= \frac{r}{R}\cdot\frac{R^2}{r^2}\cdot\frac{r}{R} \\ &= 1. \end{aligned} \tag{C.12}$$

Here, Eqs. (C.2) and (C.11) have been used in the equation expansion of Eq. (C.12).

The above shows that the number of quantum states $G$ is constant from the Sun and the Earth. On the other hand, the number of photons $N$ is also constant since there is almost a vacuum between the Sun and the Earth. Therefore, the total number of microscopically accessible states for a Bose particle, $W = {}_{N+G-1}C_N$, is kept constant, which consequently proves that the total entropy, $S == k_B \ln W$, of the solar radiation (photon ensemble) from the Sun to the Earth is unchanged from the original value. The essence of this proof is the analysis in differential form. Moreover, it is important to note that the terrestrial solar radiation at any given time is what is made up of the photons emitted by the Sun at different times mixed together, assuming that the solar radiation is in the dynamic steady state. Conclusively, it can be seen that the total entropy of solar radiation from the Sun to the Earth's upper atmosphere always remains at its original value. However, it should be noted that this entropy, which was maximum when it left the Sun's surface (blackbody radiation), is no longer maximum when it reaches the Earth's upper atmosphere.

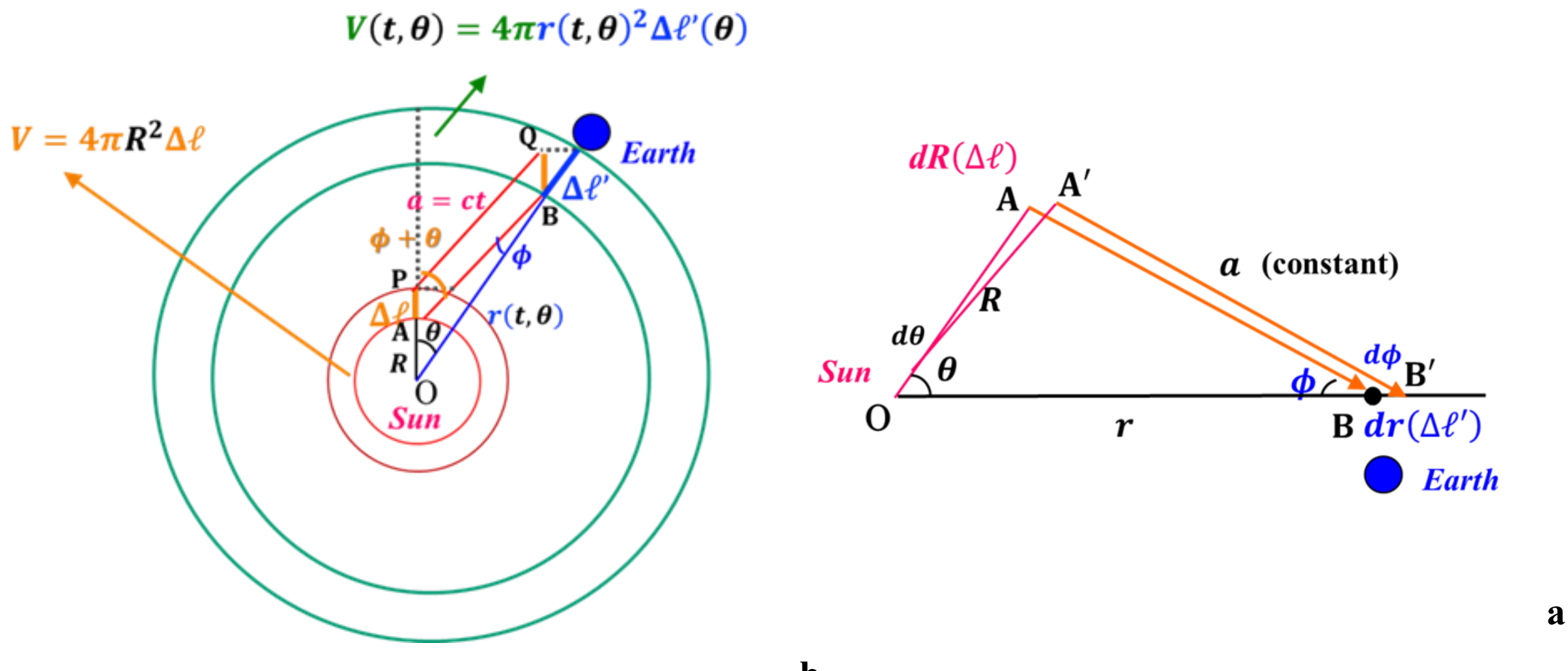

b

**Fig. 8. The cancellation between the increase in configuration-space entropy due to the dilution effect and the decrease in momentum-space entropy due to the decrease in solid angle:**

**a)** Three distances, $R$ (the radius of $the$ $S$un), $a$ (the traveling distance of a photon emitted at radiation angle $(\phi + \theta)$within time $t$, that is, $a = ct$, where $c$ is the velocity of light), $r\ (t, \theta)$(the line length OB: the distance from the center O of the Sun to the arrival point B of an emitted photon), and three angles, $\theta$ (the angle between OA and OB), $\phi$ (the zenith angle on the surface point B), and $\phi + \theta$ (the zenith angle of the emitted photon on the Sun's surface). **b)** Shell thickness $\Delta\ell = dR$ on the Sun, and corresponding spherical shell thickness $\Delta\ell'(\theta) = dr$ at the surface point B.

**Appendix D**

The formula of radiation temperature was shown in the book *Statistical Physics* (Second Revised and Enlarged Edition(1969)) written by L.D. Landau and E.M. Lifshitz [24] as follows.

$$T_{\omega,\vec{n}} = \frac{\hbar\omega}{k_B ln\left(1+\frac{\hbar\omega^3}{4\pi^3 c^3}\frac{1}{e(\omega,\vec{n})}\right)}, \tag{D.1}$$

where $e(\omega,\vec{n})d\omega do$ is the volume density of radiation in the frequency interval $d\omega$ and with the direction $\vec{n}$ of the wave vector lying in the solid-angle $do$. Performing the necessary variable transformations in Eq. (D.1) confirms that Eq. (D.1) can be attributed to the following Eq. (D.4). From the author's own description in the reference [24], it can be inferred that the following procedure was used to derive it.

In the discussion of pseudo-temperatures in Sec.4.3.2, the method of deriving the effective temperatures Eq. (4.10) was explained. The following discussion is based on this. The starting point is the following (D.2).

$$\hat{n}_{BB}(\lambda,T) = \frac{2\pi c}{\lambda^4}\frac{1}{\exp\left(\frac{hc}{\lambda k_B T}\right)-1}. \tag{D.2}$$

Let $\hat{\hat{n}}_{BB}(\lambda,T)$ be defined as the photon flux density per unit solid angle and unit wavelength interval, so that $\hat{n}_{BB}(\lambda,T) = 4\pi\,\hat{\hat{n}}_{BB}(\lambda,T)$ can be written.
As a result, Eq. (D.2) becomes

$$4\pi\,\hat{\hat{n}}_{BB}(\lambda,T) = \frac{2\pi c}{\lambda^4}\frac{1}{exp\left(\frac{hc}{\lambda k_B T}\right)-1}. \tag{D.3}$$

Substituting any $\hat{\hat{n}}(\lambda)$ in place of $\hat{\hat{n}}_{BB}(\lambda,T)$ and solving Eq. (D.3) for $T$,
we obtain

$$T\left(\lambda,\hat{\hat{n}}(\lambda)\right) = \frac{hc/\lambda}{k_B \ln\left(1+\frac{c}{\lambda 2^4 \hat{\hat{n}}(\lambda)}\right)}. \tag{D.4}$$

This Eq. (D.4) is equivalent to equation (D.1). Substituting $\hat{\hat{n}}(\lambda) = \hat{n}(\lambda)/\Omega = \hat{n}(\lambda)/(4\pi)$ into Eq.(D.4)), we obtain

$$T_\gamma\left(\lambda,\hat{\hat{n}}(\lambda)\right) = \frac{hc/\lambda}{k_B \ln\left(1+\frac{2\pi c}{\lambda^4 \hat{n}(\lambda)}\right)} \quad . \tag{D.5}$$

This formula (D.5) coincides with Eq. (4.10) in this paper. However, the temperature given by eq.(D.1) is still obtained as an extension of the blackbody radiation temperature, as the authors themselves state. Therefore, the automatic and unfounded application of this temperature to the Carnot efficiency formula to calculate the ideal efficiency of a light-powered system is not only logically unreasonable, but also does not allow the calculation of an ideal efficiency that includes the entropy reduction due to photochemical reactions within the system (in Section 6), the degree of polarization（in Section 4, and the finite absorption rate [6].